\documentclass[aps,prb,twocolumn,superscriptaddress,
groupedaddress,10pt]{revtex4}
\usepackage{amsmath}
\usepackage{amssymb}
\usepackage{graphicx}
\usepackage{epsfig}
\usepackage{dcolumn}
\usepackage{bm}
\usepackage{bm}
\usepackage{blindtext}
\usepackage{natbib}
\usepackage{siunitx} 
\usepackage{multirow}

\usepackage[titletoc]{appendix}

\usepackage{simplewick}

\usepackage{bbm}
\makeatletter
\newcommand{\mathleft}{\@fleqntrue\@mathmargin0pt}
\newcommand{\mathcenter}{\@fleqnfalse}
\makeatother

\usepackage[usenames,dvipsnames]{xcolor}
\usepackage{amsthm}
\usepackage{booktabs}
\usepackage{hyperref}
\usepackage{tikz}
\usetikzlibrary{calc}
\usepackage[printwatermark]{xwatermark}
\usepackage{mathtools}

\def\be{\begin{equation}} \def\ee{\end{equation}}
\def\bea{\begin{eqnarray}} \def\eea{\end{eqnarray}}

\def\nn{\nonumber}

\begin{document}
\title{
Origin of nonsymmorphic bosonization formulas in generalized antiferromagnetic Kitaev spin-$\frac{1}{2}$ chains
from a renormalization-group perspective
}

\author{Wang Yang}
\affiliation{Department of Physics and Astronomy and Stewart Blusson Quantum Matter Institute,
University of British Columbia, Vancouver, B.C., Canada, V6T 1Z1}

\author{Chao Xu}
\affiliation{Kavli Institute for Theoretical Sciences, University of Chinese Academy of Sciences, Beijing 100190, China}


\author{Alberto Nocera}
\affiliation{Department of Physics and Astronomy and Stewart Blusson Quantum Matter Institute, 
University of British Columbia, Vancouver, B.C., Canada, V6T 1Z1}

\author{Ian Affleck}
\affiliation{Department of Physics and Astronomy and Stewart Blusson Quantum Matter Institute, 
University of British Columbia, Vancouver, B.C., Canada, V6T 1Z1}

\begin{abstract}

Recently, in the Luttinger liquid phase of the one-dimensional generalized antiferromagnetic Kitaev spin-1/2 model,
it has been found that the abelian bosonization formulas of the local spin operators only respect the exact discrete nonsymmorphic symmetry group of the model, not the emergent U(1) symmetry.
In this work, we perform a renormalization group (RG) study to provide explanations for the origin of the U(1) breaking terms in the bosonization formulas.
We find that the lack of U(1) symmetry  originates from the wavefunction renormalization effects in the spin operators along the RG flow induced by the U(1) breaking interactions in the microscopic Hamiltonian.
In addition, the RG analysis can give predictions to the signs and order of magnitudes of the coefficients in the bosonization formulas.
Our work is helpful to understand the  rich nonsymmorphic physics in  one-dimensional Kitaev spin models.

\end{abstract}
\maketitle

\section{Introduction}

Kitaev materials have attracted intense research attentions in the past decade \cite{Jackeli2009,Chaloupka2010,Singh2010,Price2012,Singh2012,Plumb2014,Kim2015,Winter2016,Baek2017,Leahy2017,Sears2017,Wolter2017,Zheng2017,Rousochatzakis2017,Kasahara2018,Rau2014,Ran2017,Wang2017,Catuneanu2018,Gohlke2018,Liu2011,Chaloupka2013,Johnson2015,Motome2020},
since they not only provide potential experimental platforms for realizing the Kitaev spin-1/2 model on the honeycomb lattice -- a prototypical strongly correlated model for topological quantum computations \cite{Kitaev2006,Nayak2008},
but also are representatives of frustrated magnetic systems, 
having close relations to the fields of strongly correlated quantum magnetism \cite{Fazekas1999,Lauchli2006} and quantum spin liquids \cite{Balents2010,Witczak-Krempa2014,Rau2016,Winter2017,Zhou2017,Savary2017}.
Theoretical and experimental studies have established the fact that Kitaev materials can be described by 
generalized Kitaev spin models \cite{Chaloupka2010,Rau2014,Catuneanu2018,Gohlke2018,Ran2017,Wang2017} which -- in addition to Kitaev interaction -- contain other types of couplings including the Heisenberg interaction, the off-diagonal $\Gamma$ and $\Gamma^\prime$ terms, and beyond nearest neighbor interactions.
One of the central themes in the field of Kitaev materials is to understand the effects of such additional interactions which are inevitable in real materials.

Recently, there has been increasing interests in studying one-dimensional (1D) Kitaev spin models \cite{Sela2014,Agrapidis2018,Agrapidis2019,Catuneanu2019,Yang2020,Yang2020a,Yang2020b,Yang2021b,Yang2022,Yang2022b,Luo2021,Luo2021b,You2020,Sorensen2021},
which are constructed by selecting one row out of the honeycomb lattice.
These 1D Kitaev spin models have discrete nonsymmorphic symmetry group structures \cite{Yang2020,Yang2020a,Yang2022}, 
leading to rich physics including emergent conformal symmetries, extended Luttinger liquid phases in the phase diagram, nonvanishing string order parameters, and magnetically ordered phases with exotic symmetry breaking patterns such as $O_h\rightarrow D_4$ \cite{Yang2020,Yang2022b}, $O_h\rightarrow D_3$ \cite{Yang2020b,Yang2021b,Yang2022b}, $O_h\rightarrow D_2$ \cite{Yang2022b}, and $D_{3d}\rightarrow \mathbb{Z}_2$ \cite{Yang2020a}, where $O_h$ is the full octahedral group, $D_n$ is the dihedral group of order $2n$,
and $D_{3d}\cong D_3\times \mathbb{Z}_2$.
The motivation of such 1D studies is to provide hints and guidance for the 2D physics.
Indeed, it has been demonstrated in Ref. \onlinecite{Yang2022b} that the zigzag phase in 2D Kitaev-Heisenberg-Gamma model can be obtained by weakly coupling an infinite number of 1D chains, thereby providing a controllable approach to the 2D zigzag order.
In addition, 1D studies also have their independent merits, since there have been proposals on realizing  1D generalized  Kitaev spin models in real materials \cite{Motome2020}.

As shown in Ref. \onlinecite{Yang2020a}, the system has an emergent U(1) symmetry at low energies in the gapless Luttinger liquid phase in the generalized  Kitaev spin-1/2 chain with an antiferromagnetic (AFM) Kitaev coupling.
At first sight,  it seems that the discrete nature of the nonsymmorphic symmetry group is lost in the long wavelength limit.
However, as discussed in detail in Ref. \onlinecite{Yang2022b}, the discreteness of the nonsymmorphic symmetry group still has notable influence on the low energy properties,
reflected by the constraints on the abelian bosonization formulas for the spin operators.
The abelian bosonization formulas build the connections between the lattice spin operators on one side and the low energy field theory degrees of freedom on the other side, and the two sides have to be covariant under symmetry transformations. 

One  typical type of the  nonsymmorphic symmetry operations is the screw operation,
where a spatial translation followed by a spin rotation is a symmetry of the system, whereas neither the translation nor the spin rotation alone leaves the system invariant.
Unlike the on-site spin rotational symmetry in a translationally invariant system,
a screw symmetry relates the spin operators on different sites.
Hence it is expected that the constraint imposed by a screw symmetry is much looser than the constraints imposed by translation plus global spin rotation. 

Indeed, it was found in Ref. \onlinecite{Yang2022b} that the  bosonization formulas for the spin-1/2 Kitaev-Heisenberg-Gamma chain 
 contain a large number (equal to ten) of non-universal bosonization coefficients, which are only compatible with the exact nonsymmorphic symmetry group, not respecting the emergent U(1) symmetry.
The ten bosonization coefficients are  determined by density matrix renormalization group (DMRG) numerical simulations to a high degree of accuracy \cite{Yang2022b}.
However, although a symmetry analysis is able to determine the constraints on the relations among the bosonization coefficients, it cannot give any prediction on the magnitudes or signs of the coefficients, neither can it provide explanations for the mechanism of how these coefficients arise. 

In this work, in view of the aforementioned incapability of the symmetry analysis, we perform a renormalization group (RG) study in the Luttinger liquid phase of the Kitaev-Heisenberg-Gamma spin-1/2 chain in the AFM Kitaev region.
The basic idea  is that the U(1) breaking terms in the microscopic Hamiltonian renormalize the spin operators along the RG flow,
and the nonsymmorphic bosonization coefficients are reminiscences of such renormalization effects in the low energy physics.
Our RG study is able to explain the origin of the U(1)  breaking bosonization coefficients.
In addition, it can also give predictions on the signs and order of magnitudes of the bosonization coefficients.
We note that as revealed by this RG study, the U(1) breaking effects in the bosonization coefficients arise at the ``Planck scale" of the lattice,
before the lattice sites within a unit cell get smeared and lose distinguishability.
Therefore, we emphasize that our RG treatment is applied in the ultraviolet (UV) high energy region, unlike the usual cases where RG analysis is typically performed in the low energy limit.
This RG study cannot produce quantitative predictions, though indeed, it correctly captures the qualitative features of the related physics.

The rest of the paper is organized as follows.
In Sec. \ref{sec:nonsym_bosonize}, we introduce the model Hamiltonian, discuss the phase diagram of the model, and give a review on the nonsymmorphic bosonization formulas in the Luttinger liquid phase under interest.
In Sec. \ref{sec:setup_RG}, the general framework of the RG treatment in this work is formulated.
Sec. \ref{sec:derive_RG} derives and solves the RG flow equations for the scaling fields which are coupled to the spin operators.
In Sec. \ref{sec:bosonize_coeff_from_RG}, the bosonization coefficients are derived by solving the flow equations.
Finally in Sec. \ref{sec:summary}, we briefly summarize the main results of the paper.

\section{Nonsymmorphic bosonization formulas}
\label{sec:nonsym_bosonize}

\subsection{Model Hamiltonian}
\label{subsec:Ham_LL3}

\begin{figure}[h]
\begin{center}
\includegraphics[width=8.5cm]{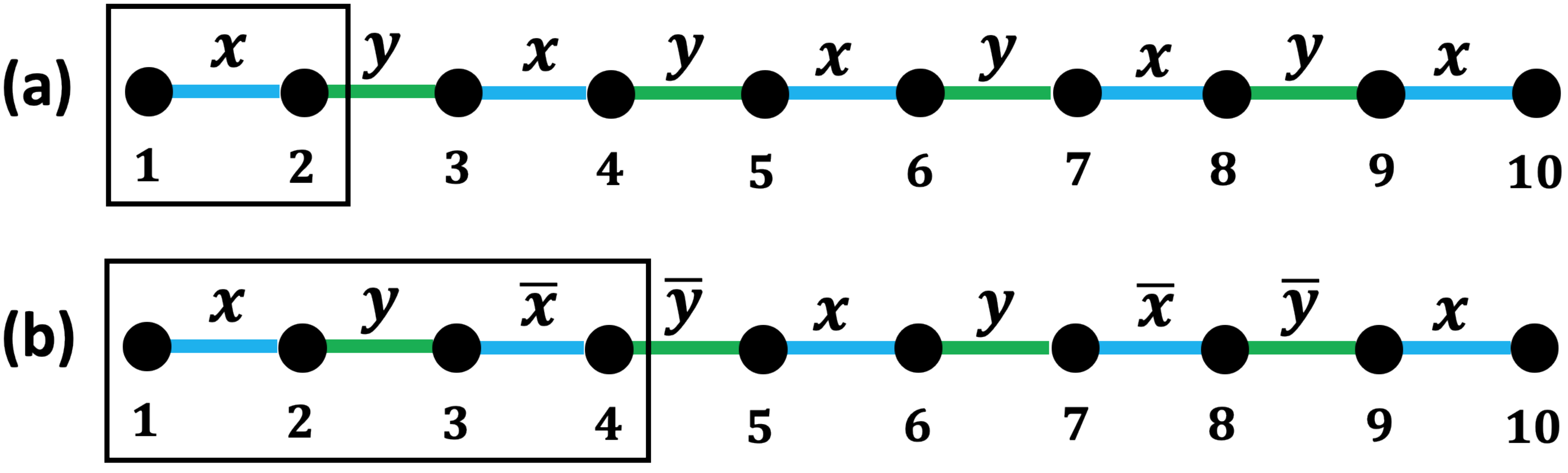}
\caption{Bond patterns of the Kitaev-Heisenberg-Gamma chain (a) before the sublattice rotation, 
and (b) after the four-sublattice rotation.
} \label{fig:bonds}
\end{center}
\end{figure}

\begin{figure}[h]
\centering
\includegraphics[width=5cm]{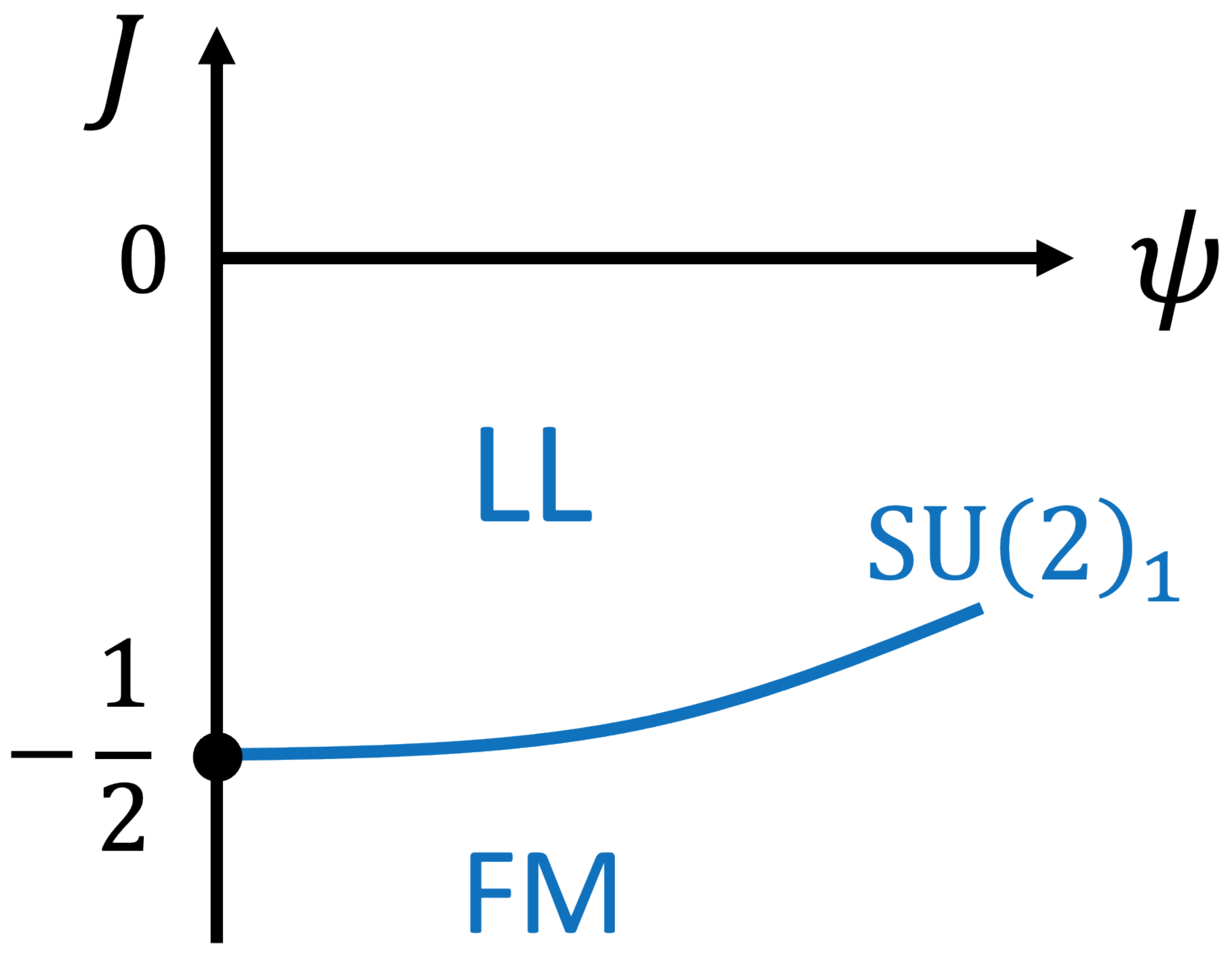}
\caption{Phase diagram of the spin-1/2 Kitaev-Heisenberg-Gamma  chain in the region $K>0,J<0$, in which the vertical axis is $J$ and the horizontal axis is $\psi$ where $K=\cos(\psi)$ and $\Gamma=\sin(\psi)$.
In the figure, ``LL" and ``FM" denote the Luttinger liquid and FM phases, respectively \onlinecite{Yang2020a,Yang2022b}.
The phase boundary between LL and FM phases is described by an emergent SU(2)$_1$ conformal symmetry at low energies \onlinecite{Yang2022b}. 
} 
\label{fig:phase}
\end{figure}

We consider a spin-1/2 Kitaev-Heisenberg-Gamma chain in zero magnetic field defined as
\begin{flalign}
H=\sum_{<ij>\in\gamma\,\text{bond}}\big[ KS_i^\gamma S_j^\gamma+ J\vec{S}_i\cdot \vec{S}_j+\Gamma (S_i^\alpha S_j^\beta+S_i^\beta S_j^\alpha)\big],
\label{eq:Ham}
\end{flalign}
in which $i,j$ are two sites of nearest neighbors;
$\gamma=x,y$ is the spin direction associated with the $\gamma$ bond shown in Fig. \ref{fig:bonds} (a);
$\alpha\neq\beta$ are the two remaining spin directions other than $\gamma$;
$K$, $J$ and $\Gamma$,
are the Kitaev, Heisenberg and Gamma couplings, respectively.
The coupling constants $K,\Gamma$ can be parametrized as 
 $K=\cos(\psi)$, $\Gamma=\sin(\psi)$, in which $\psi\in[0,\pi]$.
The phase diagram of the model in terms of $J,\psi$ is shown in Fig.\ref{fig:phase}.

A useful unitary transformation is called four-sublattice rotation $U_4$, which is defined as
\begin{eqnarray}
\text{Sublattice $1$}: & (x,y,z) & \rightarrow (-x^{\prime},y^{\prime},-z^{\prime}),\nn\\ 
\text{Sublattice $2$}: & (x,y,z) & \rightarrow (-x^{\prime},-y^{\prime},z^{\prime}),\nn\\
\text{Sublattice $3$}: & (x,y,z) & \rightarrow (x^{\prime},-y^{\prime},-z^{\prime}),\nn\\
\text{Sublattice $4$}: & (x,y,z) & \rightarrow (x^{\prime},y^{\prime},z^{\prime}),
\label{eq:4rotation}
\end{eqnarray}
in which ``Sublattice $i$" ($1\leq i \leq 4$) represents all the sites $i+4n$ ($n\in \mathbb{Z}$) in the chain, and we have dropped the spin symbol $S$ for simplicity (i.e., $\alpha$ is understood as $S^\alpha$ where $\alpha=x,y,z$).
The Hamiltonian $H^{\prime}=U_4 H U_4^{-1}$ in the four-sublattice rotated frame acquires the form 
\begin{eqnarray}
H^{\prime}&=&\sum_{<ij>\in \gamma\,\text{bond}}\big[ (K+2J)S_i^\gamma S_j^\gamma-J \vec{S}_i\cdot \vec{S}_j \nn\\
&&+\epsilon(\gamma) \Gamma (S_i^\alpha S_j^\beta+S_i^\beta S_j^\alpha)\big],
\label{eq:4rotated}
\end{eqnarray}
in which the bonds $\gamma=x,y,\bar{x},\bar{y}$ has a four-site periodicity as shown in Fig. \ref{fig:bonds} (b);
the function $\epsilon(\gamma)$ is defined as $\epsilon(x)=\epsilon(y)=-\epsilon(\bar{x})=-\epsilon(\bar{y})=1$;
$S_i^{\bar{\gamma}}=S_i^{\gamma}$;
and $\vec{S}_i^{\prime}=U_4\vec{S}_iU_4^{-1}$ is denoted as $\vec{S}_i$ for short.
Notice that $U_4$ reveals a hidden SU(2) symmetric point located at $K+2J=0$, $\Gamma=0$.
At this point, $H^\prime$ is exactly the SU(2) symmetric AFM Heisenberg model. 
Explicit forms of $H$ and $H^\prime$ are included in Appendix \ref{app:Ham}.

From here on, we will stick to the four-sublattice rotated frame unless otherwise stated.

\subsection{Phase diagram in the antiferromagnetic Kitaev region}
\label{subsec:phase_diagram_LL3}

The phase diagram in the region $K>0$, $J<0$ is shown in Fig. \ref{fig:phase}.
Since a global spin rotation around $z$-axis by $\pi$ changes the sign of $\Gamma$ but leaves $K$ and $J$ invariant,
it is enough to consider the $\Gamma>0$ region.

As can be seen from Fig. \ref{fig:phase}, there are two phases close to the $\Gamma=0$ line (i.e., the vertical axis), including a Luttinger liquid phase (denoted as LL in Fig. \ref{fig:phase}), and a ferromagnetically ordered phase (denoted as FM).
It has been shown in Ref. \onlinecite{Yang2022b} that in the sense of low energy field theory, the phase boundary between the LL and FM is essentially a phase transition between planar and axial spin-1/2 XXZ chains.
Hence, the low energy physics of this phase boundary is described by the SU(2)$_1$ Wess-Zumino-Witten (WZW) model.

In this paper, we will focus on the Luttinger liquid phase in Fig. \ref{fig:phase}.

\subsection{Nonsymmorphic abelian bosonization formulas}
\label{subsec:bosonization_LL3}

In this subsection, we briefly review the nonsymmorphic bosonization formulas in the Luttinger liquid phase in Fig. \ref{fig:phase},
which are proposed  in Ref. \onlinecite{Yang2022b} based on a symmetry analysis.

The system in the four-sublattice rotated frame is invariant under the following symmetry operations \onlinecite{Yang2020a,Yang2022b},
\begin{eqnarray}
T &: & (S_i^x,S_i^y,S_i^z)\rightarrow (-S_{i}^x,-S_{i}^y,-S_{i}^z)\nn\\
R(\hat{y},\pi)I&: & (S_i^x,S_i^y,S_i^z)\rightarrow (-S_{5-i}^x,S_{5-i}^y,-S_{5-i}^z)\nn\\
R(\hat{z},-\frac{\pi}{2})T_a&:& (S_i^x,S_i^y,S_i^z)\rightarrow (-S_{i+1}^y,S_{i+1}^x,S_{i+1}^z),
\label{eq:symmetries}
\end{eqnarray}
in which $T$ is time reversal; $I$ is the spatial inversion with inversion center located at the middle of the bond connecting sites $2$ and $3$; $T_{na}$ is the spatial translation by $n$ sites; and $R(\hat{n},\theta)$ represents a global spin rotation around $\hat{n}$-axis by an angle $\theta$.
It has been proved in Refs. \onlinecite{Yang2020a,Yang2022b} that the symmetry group $G=\mathopen{<}T,R(\hat{y},\pi)I,R(\hat{z},-\frac{\pi}{2})T_a\mathclose{>}$ is nonsymmorphic and satisfies $G/\mathopen{<}T_{4a}\mathclose{>}\cong D_{4d}$,
in which $\mathopen{<}...\mathclose{>}$ represents the group generated by the elements within the bracket; 
and $D_{4d}\cong \mathbb{Z}_2\times D_4$.

In the Luttinger liquid phase, the low energy theory is described by the Luttinger liquid Hamiltonian
\bea
H_{LL}=\frac{v}{2} \int dx [\kappa^{-1} (\nabla \varphi)^2 +\kappa (\nabla \theta)^2],
\label{eq:H_LL}
\eea
in which $v$ is the velocity; $\kappa$ is the Luttinger parameter; and the fields $\theta,\varphi$ satisfy $[\varphi(x),\theta(x^\prime)]=\frac{i}{2}\text{sgn}(x^\prime-x)$.
For later convenience, it is useful to define the following fields,
\begin{flalign}
&J^\pm=\frac{2}{a}\cos(\sqrt{4\pi}\varphi)e^{\pm i\sqrt{\pi}\theta},~  J^z=-\sqrt{2}\pi \nabla \varphi,\nn\\ 
&N^\pm=\frac{\sqrt{2}}{a} e^{\pm i \sqrt{\pi}\theta},~   N^z=\frac{\sqrt{2}}{a}\sin(\sqrt{4\pi}\varphi), 
\end{flalign}
where $J^\pm=J^x\pm iJ^y$ and $N^\pm=N^x\pm iN^y$ \onlinecite{}.
Since $\int dx J^z(x)$ is the generator for the global spin rotation around $z$-axis, 
$J^\alpha$ and $N^\alpha$ transform under $R(\hat{z},\beta)$ as $A^\pm\rightarrow A^\pm e^{\pm i\beta}$ 
and $A^z\rightarrow A^z$, where $A=J,N$. 
Clearly, the low energy field theory has an emergent U(1) symmetry corresponding to rotations around $z$-axis,
even though the microscopic Hamiltonian only has a discrete nonsymmorphic symmetry group. 

On the other hand, the discrete and nonsymmorphic nature of the symmetry group still has significant effects on the low energy properties of the system.
We note that when the microscopic Hamiltonian is U(1) invariant (for example, the planar XXZ model), the bosonization formulas of the spin operators are given by 
$S_j^\alpha=\lambda J^\alpha+\mu(-)^j  N^\alpha$, in which $\lambda,\mu$ are constants. 
However, these relations cease to apply in the Kitaev-Heisenberg-Gamma chain.
In Ref. \onlinecite{Yang2022b}, 
the following nonsymmorphic bosonization formulas are proposed
\begin{eqnarray}
S^{\alpha}_{j+4n} = \sum_\beta [D_{j}^{\alpha\beta} J^{\beta}(x)+(-)^j C_{j}^{\alpha\beta} N^{\beta}(x)],
\label{eq:nonsym_bosonization}
\end{eqnarray}
in which: $n$ is the index for the unit cell; $j$ ($1\leq j\leq 4$) represents the site within the four-site unit cell; $x=j+4n$ is the spatial coordinate in the continuum limit;
and $\alpha,\beta=x,y,z$.

Two comments are in order.
First, Eq. (\ref{eq:nonsym_bosonization}) was obtained in Ref. \onlinecite{Yang2022b}
by covariance of the two sides under symmetry transformations.
Notice that for non-symmetry transformations, the two sides in Eq. (\ref{eq:nonsym_bosonization}) are not covariant,
since the transformed $J^{\beta}$ and $N^{\beta}$ operators are driven out of the low energy subspace of the Hilbert space in such situations. 
Second, Eq. (\ref{eq:nonsym_bosonization}) equally applies to nonabelian bosonization formulas in the nonsymmorphic case,
except that the $J^{\beta}$ and $N^{\beta}$ operators should be replaced by the WZW current operators and primary fields, respectively. 
As shown in Fig. \ref{fig:phase}, the line separating LL and FM phases has an emergent SU(2)$_1$ conformal  symmetry at low energies (see Ref. \onlinecite{Yang2022b} for details). 
Therefore, a nonabelian bosonization version of Eq. (\ref{eq:nonsym_bosonization}) should be used along this phase transition line. 

Defining $3\times 3$ matrices $D_{j}$ and $C_{j}$ whose matrix elements at position $(\alpha,\beta)$ are $D_{j}^{\alpha\beta}$ and $C_{j}^{\alpha\beta}$,
the coefficients in Eq. (\ref{eq:nonsym_bosonization}) can be compactly expressed as
\bea
D_1&=&\left(\begin{array}{ccc}
a_D & b_D & c_D\\
b_D & a_D & -c_D\\
h_D & -h_D & i_D
\end{array}\right), \nn\\
D_j&=&(M_z)^{j-1} D_1 (M_z)^{1-j},\nn\\
\label{eq:coeff_matrix_D}
\eea
and
\begin{eqnarray}
C_1&=&\left(\begin{array}{ccc}
a_C & b_C & c_C\\
b_C & a_C & -c_C\\
h_C & -h_C & i_C
\end{array}\right), \nn\\
C_j&=&(M_z)^{j-1} C_1 (M_z)^{1-j},\nn\\
\label{eq:coeff_matrix_C}
\end{eqnarray}
where $j=2,3,4$, and 
\bea
M_z=\left(
\begin{array}{ccc}
0&1&0\\
-1&0&0\\
0&0&1
\end{array}
\right).
\eea
It can be seen that there are ten non-universal coefficients in Eq. (\ref{eq:nonsym_bosonization}), which spoil the emergent  U(1) symmetry and only respect the exact nonsymmorphic symmetries of the system.
Explicit expressions of the nonsymmorphic bosonization formulas are included in Appendix \ref{app:bosonization}.

On the other hand, although the symmetry analysis is able to determine the form of the bosonization formulas, it has no predictive power on the order of magnitudes nor the signs of the ten bosonization coefficients $a_\Lambda,b_\Lambda,c_\Lambda,h_\Lambda,i_\Lambda$ ($\Lambda=C,D$).
In addition, the symmetry analysis gives no explanation to the origin of the bosonization coefficients, i.e., there is no information on how they arise microscopically. 
In view of these issues, it is the purpose of this work to derive the ten bosonization coefficients using an RG approach.

\section{Setup for RG flows}
\label{sec:setup_RG}

In this section, we set up the method for deriving the RG flow equations which can provide explanations for the microscopic origin of the nonsymmorphic  bosonization coefficients.

The low energy physics of the 1D spin-1/2 repulsive Hubbard model at half filling is known to be described by the SU(2)$_1$ Wess-Zumino-Witten (WZW) model, which is the same as the low energy theory of the spin-1/2 AFM Heisenberg model (for details, see \onlinecite{Affleck1988} and Appendix \ref{app:nonabelian_bosonization}).
Hence, the weak coupling repulsive Hubbard model can be used to mimic the low energy physics of the Kitaev-Heisenberg-Gamma model at the hidden AFM point (i.e., $K+2J=0$, $\Gamma=0$) in the four-sublattice rotated frame.
Then $K+2J$ and $\Gamma$ can be treated as perturbations to the repulsive Hubbard model. 

Here we make some comments  on the reasons why a fermion model has to be introduced for an RG treatment, and the limitations of the method.
We first emphasize that the bosonization coefficients arise from the microscopic lattice structures.
Hence a perturbation in the low energy sector cannot capture these bosonization coefficients, and the physics at the ``Planck scale" of the lattice has to be involved.
It seems that there is still hope since the spin-1/2 Heisenberg model is an integrable system solvable by  the Bethe ansatz method, 
which is applicable to any energy scale.
However, a perturbation on the Heisenberg model is analytically intractable since Bethe ansatz is a very intricate method, not suitable for perturbative calculations.

On the other hand, it is standard to perform perturbative calculations based on the free fermion models.
Therefore, in the weak coupling limits, i.e., when the Hubbard interaction, the combination $K+2J$, and the Gamma interaction are all small,
an RG analysis can be applied in the vicinity of the free fermion fixed point.
Notice that this directly implies the limitation of the method.
Our RG analysis is only qualitative, since the model is changed from a pure spin model to a fermion model.
However, this RG analysis is able to provide explanations for the origin of the bosonization coefficents, justifying the proposed nonsymmorphic bosonization formulas in Eq. (\ref{eq:nonsym_bosonization}).
It is able to give predictions on the signs and order of magnitudes of the bosonization coefficients.

We start from the following fermion model in the four-sublattice rotated frame,
\bea
H_F&=&H_0+H_{int},
\eea
in which
\bea
H_0&=&-t\sum_{<ij>,\alpha} (c_{i\alpha}^\dagger c_{j\alpha}+\text{h.c.})-\mu\sum_{i\alpha}c_{i\alpha}^\dagger c_{i\alpha},\nn\\
H_{int}&=&H_U+H_4,
\label{eq:fermion_Ham}
\eea
where 
\begin{flalign}
&H_U=U\sum_i n_{i\uparrow}n_{i\downarrow}\nn\\
&H_4=\sum_{<ij>\in \gamma\,\text{bond}}\big[ (K+2J)S_i^\gamma S_j^\gamma+\epsilon(\gamma) \Gamma (S_i^\alpha S_j^\beta+S_i^\beta S_j^\alpha)\big].
\end{flalign}
At half-filling and for a repulsive $U$, $H_0+H_U$ reproduces  the SU(2)$_1$ WZW model in the low energy limit (see Ref. \onlinecite{Affleck1988} and Appendix. \ref{app:nonabelian_bosonization}).
Then by adding $H_4$, the low energy physics of the Kitaev-Heisenberg-Gamma model is recovered.

The partition function for $H_F$ is given by
\bea
\mathcal{Z}=\int D[c,c^\dagger] e^{-\mathcal{S}},
\eea
where 
\bea
\mathcal{S}=\int d\tau(\sum_{i\alpha} c^\dagger_{i,\alpha} \partial_\tau c_{i\alpha} +H_0+H_{int}).
\label{eq:action}
\eea
The goal is to compute the spin correlation functions $G_{i\alpha,j\beta}$
\bea
G_{i\alpha,j\beta}(\tau,n)&=&\langle S_i^\alpha(0) S_{j+4n}^\beta(\tau)\rangle\nn\\
&=&\frac{1}{\mathcal{Z}}\int D[c,c^\dagger]S_i^\alpha(0) S_{j+4n}^\beta(\tau)  e^{-\mathcal{S}}
\eea
in which $\tau$ is the imaginary time, $i,j\in\{1,2,3,4\}$, and 
\bea
S_{i}^\alpha=\sum_{a,b}\frac{1}{2}c_{ia}^\dagger \sigma^\alpha_{ab} c_{ib},
\eea
where $a,b=\uparrow,\downarrow$.

\begin{figure}[h]
\begin{center}
\includegraphics[width=4cm]{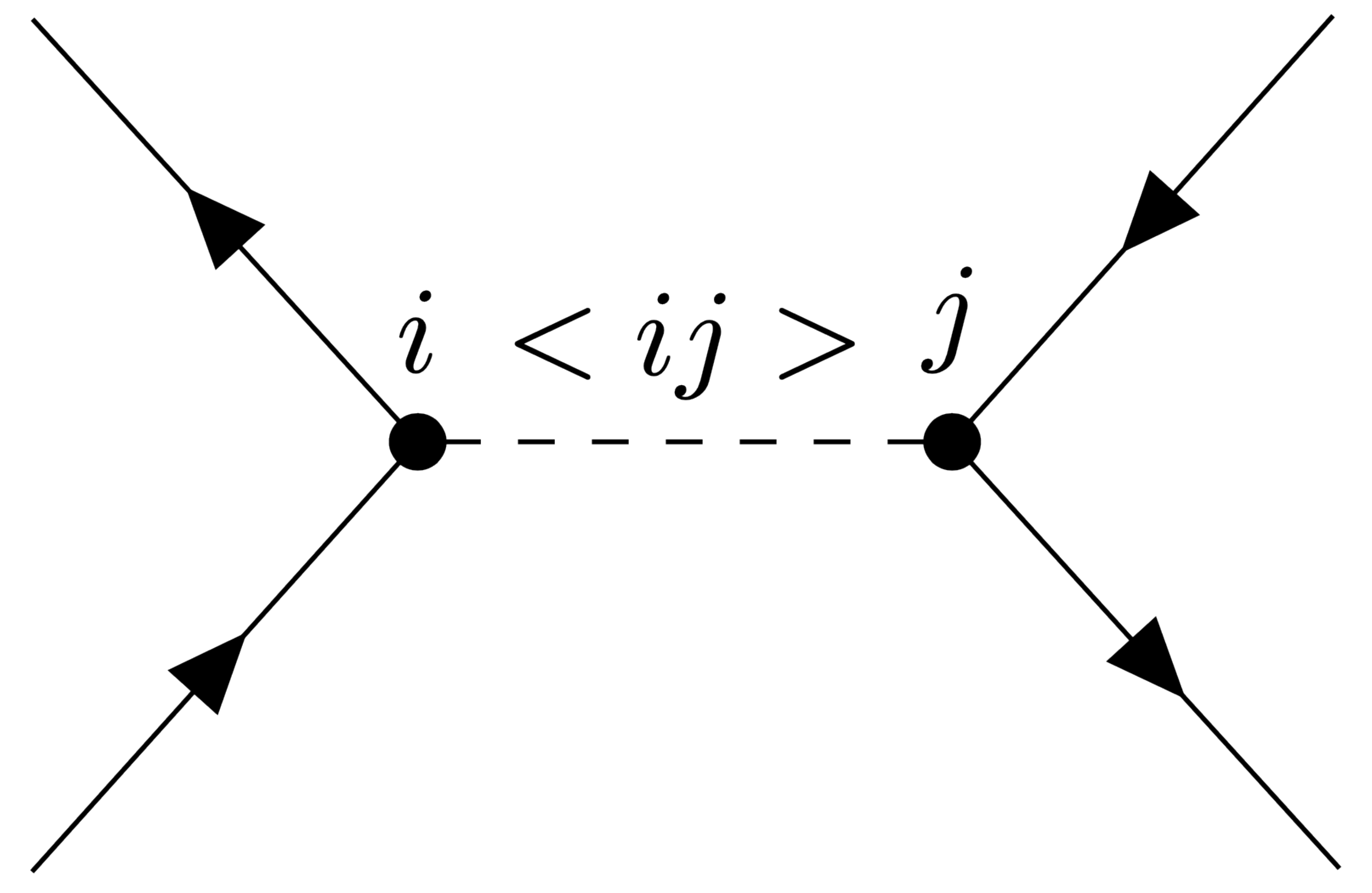}
\caption{Diagrammatic representation of the interaction term between site $i$ and $j$, where $1\leq i,j\leq 4$.
The outgoing arrow, ingoing arrow, and dashed lines represent fermion creation operator, fermion annihilation operator, and the four-fermion interaction, respectively. 
} \label{fig:diagram_int}
\end{center}
\end{figure}

\begin{figure}[h]
\begin{center}
\includegraphics[width=3cm]{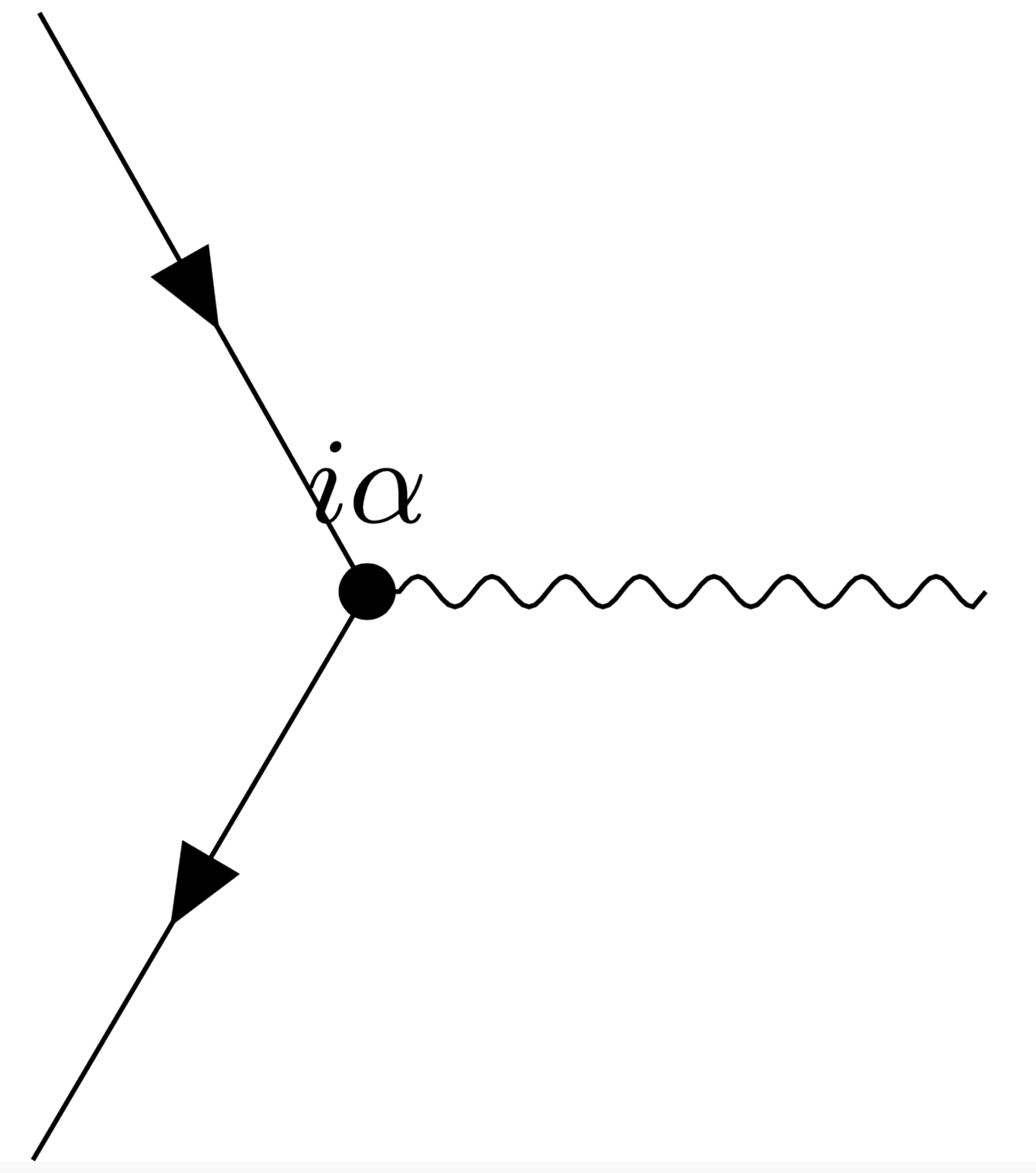}
\caption{Diagrammatic representation of the spin operator.
The outgoing arrow, ingoing arrow, and wavy lines represent fermion creation operator, fermion annihilation operator, and external magnetic field, respectively. 
} \label{fig:diagram_spin}
\end{center}
\end{figure}

We note that upon integrating over the fast modes in a momentum shell, $H_{int}$ renormalizes the spin operators.
In fact, by separating the fast and slow modes, we obtain
\begin{flalign}
&G_{i\alpha,j\beta}(\tau,n)=\frac{1}{\mathcal{Z}_<}\int D[c_<,c^\dagger_<] e^{-\mathcal{S}_{eff,<}}
 \nn\\
& \langle(1-\int d\tau dx H_{int,>,<})S_i^\alpha(0) S_{j+4n}^\beta(\tau)\rangle_>,
\label{eq:renormalize_interaction}
\end{flalign}
in which $H_{int,>,<}$ represents the mixing term between the fast and slow modes, $\langle...\rangle_>$ is defined as
\bea
\langle...\rangle_>=\frac{1}{\mathcal{Z}_>}\int D[c_>,c^\dagger_>] e^{-\mathcal{S}_{eff,>}}(...),
\eea
and only first order renormalization is taken into account for the spin operators. 
Eq. (\ref{eq:renormalize_interaction}) leads to a set of coupled Callan-Symanzik equations \cite{Amit1984}, which can be solved to determine the behaviors of the correlation functions.

\begin{figure}[h]
\begin{center}
\includegraphics[width=6cm]{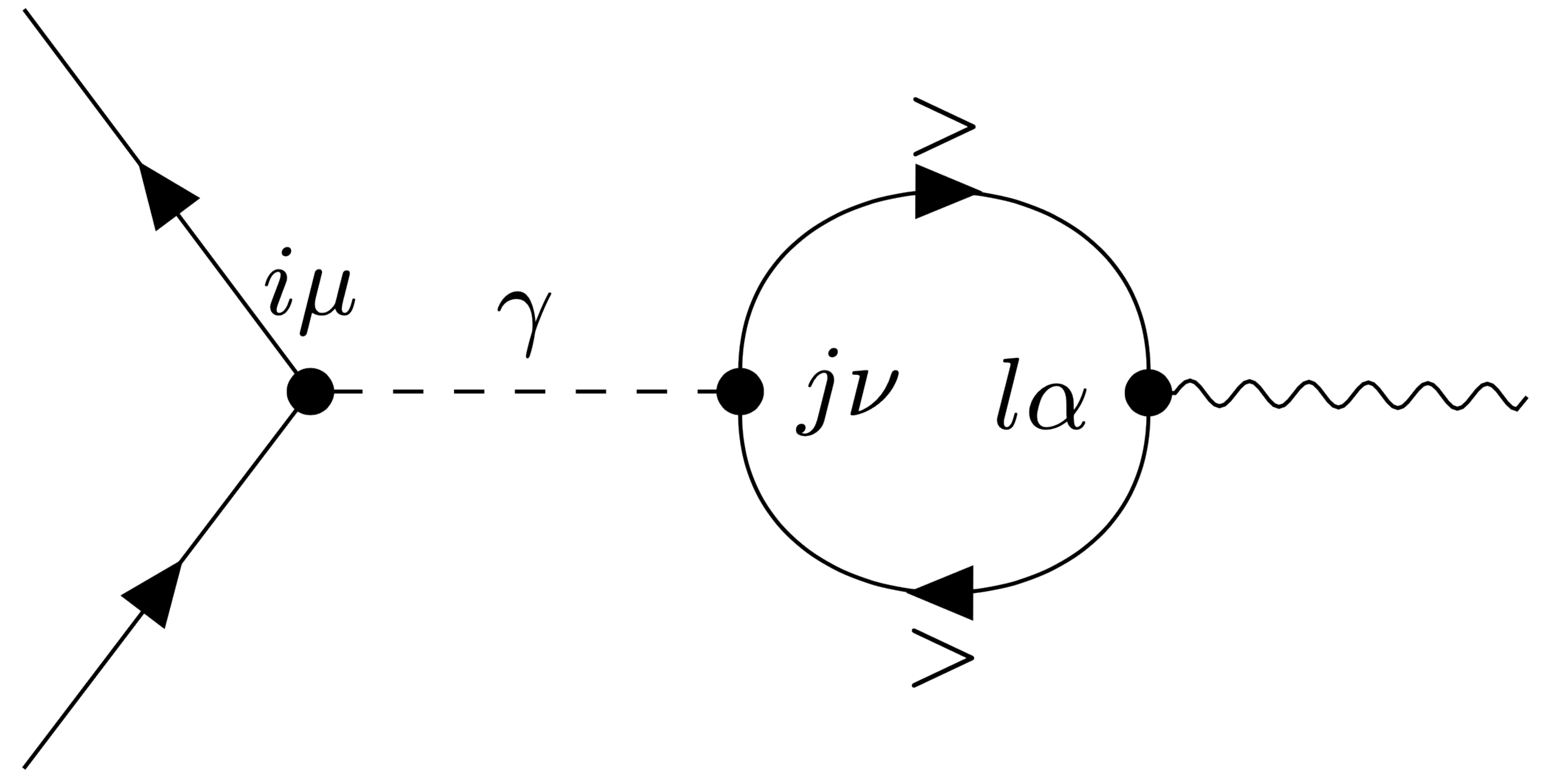}
\caption{Diagrams for the renormalization of the spin operators.
} \label{fig:renormalize1}
\end{center}
\end{figure}

\begin{figure}[h]
\begin{center}
\includegraphics[width=4.5cm]{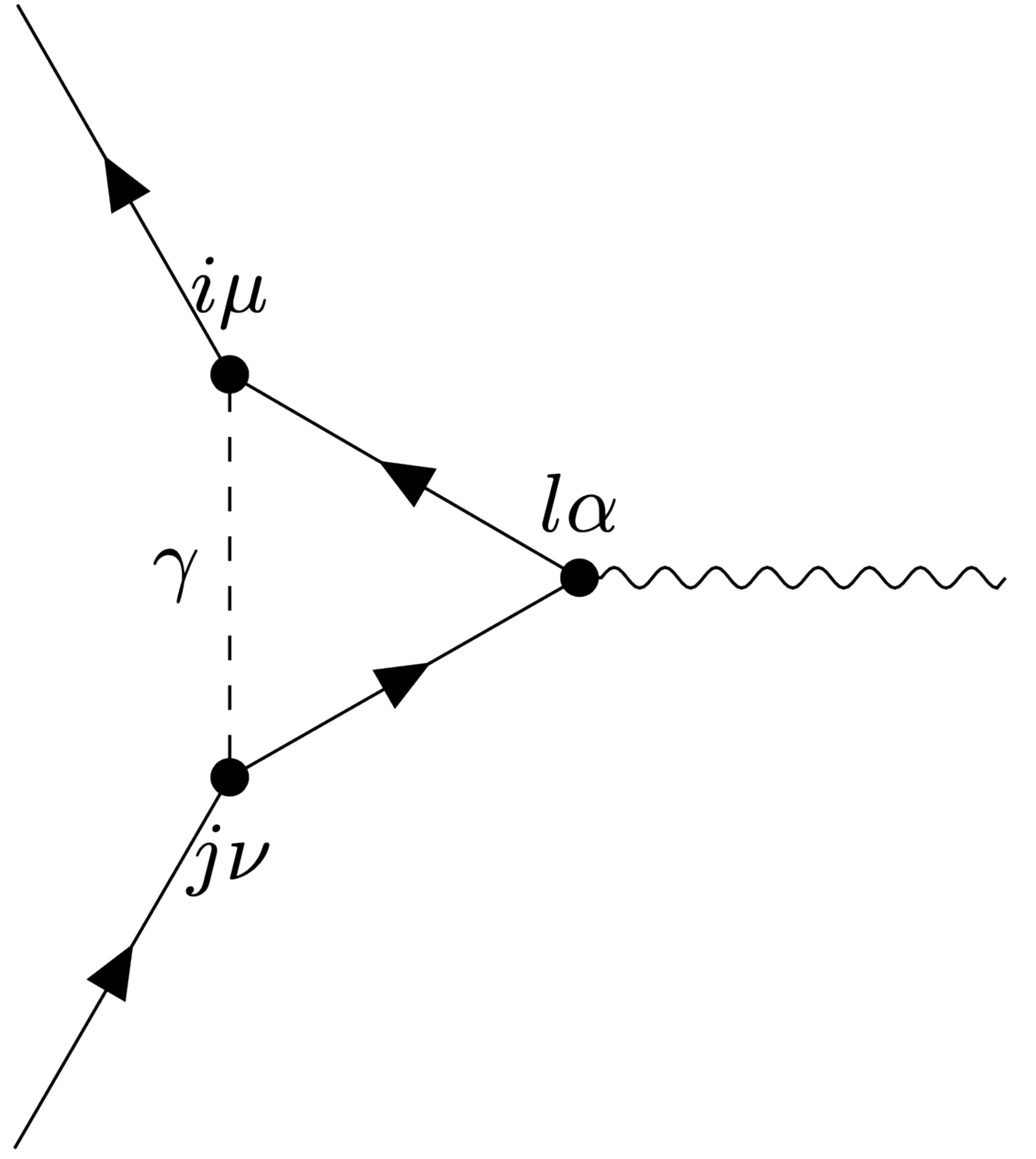}
\caption{Diagrams that generate new terms upon RG flow.
} \label{fig:renormalize2}
\end{center}
\end{figure}

The interactions and the spin operators are represented by the diagrams in Fig. \ref{fig:diagram_int} and Fig. \ref{fig:diagram_spin}, respectively. 
In particular, Fig. \ref{fig:diagram_int} represents $(K+2J)S_i^\gamma S_j^\gamma +\epsilon(\gamma) \Gamma (S_i^\alpha S_j^\beta+S_i^\beta S_j^\alpha)$ where $\mathopen{<}ij\mathclose{>}=\gamma$.
There are two diagrams which contribute to the contractions between $H_{int,>,<}$ and the spin operators as shown in Fig. \ref{fig:renormalize1} and Fig. \ref{fig:renormalize2}.
It is clear that Fig. \ref{fig:renormalize1} introduces a renormalization of the spin operators,
whereas on the other hand, Fig. \ref{fig:renormalize2} produces new terms along the RG flow, which are of the forms $c^\dagger_i \sigma^\lambda c_{j}$ where $i =j\pm 1$.
Although $c^\dagger_i \sigma^\lambda c_{j}$ is not of the form of an on-site spin operator,  it becomes indistinguishable from a spin operator in the low energy limit when the difference between adjacent sites is smeared out.
Later in Sec. \ref{sec:diagram_lambda3} we will see that the value of the diagram in Fig. \ref{fig:renormalize2} vanishes.
However, we still include it here from a conceptual consideration,  
and in addition, if the Hamiltonian in Eq. (\ref{eq:Ham}) contains beyond next-nearest neighbor terms (which is always the case in real materials), the diagram in Fig. \ref{fig:renormalize2} indeed contributes. 
We note that the set of coupled Callan-Symanzik equations for the correlation functions to the one-loop level can be obtained from the two diagrams in Fig. \ref{fig:renormalize1} and Fig. \ref{fig:renormalize2}.

Here we take an alternative route for later convenience.
Instead of considering the Callan-Symanzik equations, we introduce the following set of magnetic fields into the action
\begin{flalign}
&-\int d\tau \big[\sum_n\sum_{j=1}^4\sum_{\alpha=x,y,z} h_j^\alpha (\tau,n) S_{j+4n}^\alpha(\tau)\nn\\
&+\sum_n\sum_{i=1}^4\sum_{j=i\pm 1}\sum_{\alpha=x,y,z} \frac{1}{2} h_{ij}^\alpha (\tau,n) c^\dagger_{i+4n}(\tau) \sigma^\alpha c_{j+4n}(\tau)\big],
\end{flalign}
in which $n$ is the index of the unit cell;,$i$ and $j$ are site indices within a unit cell, and the $h_{ij}^\alpha (\tau,n) $ terms are inserted since they can be generated upon RG flow as a result of the diagram in  Fig. \ref{fig:renormalize2}.
The  spin correlation functions can be obtained from the functional derivatives as
\bea
\langle S_i^\alpha(\tau,n) S_j^\beta(\tau^\prime,n^\prime) \rangle=\frac{\partial^2 F}{\partial h_i^\alpha(\tau,n) \partial h_j^\beta(\tau^\prime,n^\prime)},
\eea
where $F=-\ln \mathcal{Z}$ is the free energy.

We will determine the RG flow equations for the scaling fields $h_j^\alpha (\tau,n)$ and $h_{ij}^\alpha (\tau,n)$.
In Eq. (\ref{eq:fermion_Ham}), 
the free fermion Hamiltonian $H_0$ is gapless at $\pm k_F=\pm \pi/(2a)$,  
giving rise to left mover $c_{L a}$ and right mover $c_{R a}$ ($a=\uparrow,\downarrow$) at low energies,
where $c_{L a}$ and $c_{R a}$ are the fermion annihilation operators for the left and right movers, containing Fourier components with wavevectors close to $-k_F$ and $k_F$, respectively. 
Then in the low energy limit, the wavevectors in the spin operators are either close to zero or $\pi$,
corresponding to intra-mover and inter-mover contributions. 
Keeping only the low energy modes,
the spin operator $\tilde{S}^\alpha_r(\tau)$ at smeared position $r$ and time $\tau$ can be written as
\bea
\tilde{S}^\alpha_r(\tau)\sim S^\alpha_u(\tau,r)+(-)^r S^\alpha_s(\tau,r),
\label{eq:def_S_u_s}
\eea
in which the uniform component $S^\alpha_u(\tau,r)$ and staggered component $S^\alpha_s(\tau,r)$ are given by
\bea
S_u^\alpha &=& \frac{1}{2}(c_L^\dagger \sigma^\alpha c_L+c_R^\dagger \sigma^\alpha c_R),\nn\\
S_s^\alpha &=& \frac{1}{2}(c_L^\dagger \sigma^\alpha c_R+c_R^\dagger \sigma^\alpha c_L),
\label{eq:S_us_c}
\eea
where $c_\lambda=(c_{\lambda\uparrow},c_{\lambda\downarrow})^T$ ($\lambda=L,R$)
and both $S^\alpha_u$ and $S^\alpha_s$ are smooth functions of $r$ (i.e., no Fourier components with a wavevector far from zero).
Here we note that since $h_l^\alpha(\tau,n)$ is defined every four sites,  the zero- and $\pi$-wavevector components cannot be distinguished in $h_l^\alpha(\tau,n)$ or $h_{ij}^\alpha (\tau,n)$ since both  components are smooth in $n$.


Finally we make a comment on the energy scales in the problem.
There are five characteristic energy scales $\Lambda_0$, $\Lambda_s$, $\Lambda_L$, $m_c$, and $E$,
where $\Lambda_0\sim 1/a$ is the UV cutoff of the lattice structure,
$\Lambda_s\sim 1/(4a)$ is the energy scale where the four sites within a unit cell are smeared and can no longer be clearly distinguished,
$\Lambda_L$ is the energy scale where a linearization of the free fermion spectrum around $\pm k_F$ can be performed,
$m_c\sim e^{-\text{const.}t/U}$ is the charge gap due to the repulsive Hubbard term,
and  $E$ is the energy scale of  the correlation functions which we are eventually  interested in.
The hierarchy of the energy scales is  clearly
\bea
\Lambda_0 \gg \Lambda_s \gg \Lambda_L\gg m_c\gg E.
\label{eq:energy_hierarchy}
\eea
We note that below  $\Lambda_L$,  the fermion has an emergent Lorentz symmetry, and  is fractionalized into a U(1) charge boson and an SU(2)$_1$ spin boson \cite{Affleck1988}.
When the energy is further lowered below $m_c$, the charge boson is gapped, and we are left with only spin degrees of freedom $J^\alpha$ and $N^\alpha$.
We also note that since the microscopic lattice structure is lost at $\Lambda_s$,
our RG analysis  stops at an energy scale $\sim\Lambda_s$.

\section{RG flow equations}
\label{sec:derive_RG}

In this section, we derive the RG flow equations for the scaling fields $h_l^\alpha(\tau,n)$ and $h_{ij}^\alpha (\tau,n)$
from the diagrams in Fig. \ref{fig:renormalize1} and Fig. \ref{fig:renormalize2}.


\subsection{Flow equations from the diagram in Fig. \ref{fig:renormalize1}}

Let's first consider the diagram in Fig. \ref{fig:renormalize1}.
It is nonvanishing when $\nu=\alpha$, and renormalizes $h_i^\mu$.
Suppose we lower the cutoff from $\Lambda_0/b$ to $\Lambda_0/b^\prime$,
where $\Lambda_0$ is chosen as $\pi/(2a)$ and $b^\prime=b+\Delta b$ with $0<\Delta b\ll 1$.
The perturbation process in Fig. \ref{fig:renormalize1} gives rise to the following term in the action
\bea
\lambda_{jl} \delta_{\nu\alpha} \Delta\ln b\int d\tau \sum_n h_l^\alpha(\tau,n) S_{i+4n}^\mu(\tau),
\label{eq:renormalize_diagram1}
\eea
leading to a renormalization of $h_i^\mu$ by $h_l^\alpha$,
where $\Delta \ln b =\Delta b/b$.
Define the free fermion Green's function $\mathcal{G}(k)$ as
\bea
\mathcal{G}(k)=\frac{1}{i\omega-\epsilon(k)}. 
\label{eq:free_f_green2}
\eea
in which $k=(i\omega,\vec{k})$ where $\omega$ is Matsubara frequency and $\vec{k}$ is the wavevector in space (we define the spatial wavevector as a vector to distinguish it from the spacetime combined index $k$, even though the system is 1D and $\vec{k}$ is in essence a scalar), 
and $\epsilon(k)$ is the free fermion dispersion which includes the chemical potential term. 
The coefficient $\lambda_{jl}$ can be derived as
\begin{flalign}
&\lambda_{jl} \Delta\ln b =  \nn\\
&-\frac{a}{8}  \sum_{m=1}^4 e^{-i\frac{\pi}{2}m(j-l)}\int_{\Lambda_0/b^\prime}^{\Lambda_0/b} d^2k  \mathcal{G}(k) \mathcal{G}(k+\frac{\pi }{2a} m \hat{x}),
\label{eq:lambda_coefficient2}
\end{flalign}
where $a$ is the lattice spacing, and  $\hat{x}$ is the unit vector in the spatial direction.
We note that because of the translation and inversion  symmetries of the free fermion theory,
$\lambda_{ij}$ satisfies the following relations
\bea
\lambda_{ij}=\lambda_{i+l,j+l}=\lambda_{-i,-j}=\lambda_{i,j+4},
\label{eq:lambda_prime}
\eea
where $l\in \mathbb{Z}$.

We briefly describe the derivation of $\lambda_{jl}$.
Detailed derivations are included in Appendix \ref{app:diagram1}.
The Fourier transforms of the fermion operator and the scaling field are defined as
\bea
 c^\dagger(k)=\frac{1}{\sqrt{N\beta}}\int d\tau \sum_{j=1}^N  c^\dagger_j(\tau) e^{i(\omega\tau-\vec{k}\cdot ja\hat{x})},
 \label{eq:fourier_c}
\eea
and
\bea
h_l^\alpha(q)=\frac{1}{\sqrt{N\beta}}\int d\tau \sum_{n=1}^{N/4} h_l^\alpha(\tau,n) e^{i(\omega\tau-\vec{q}\cdot4na\hat{x})},
\label{eq:fourier_h}
\eea
in which $N$ is the system size, $\beta$ is the inverse of the temperature,
$j$ in Eq. (\ref{eq:fourier_c}) is summed over all sites in the chain,
and $n$ in Eq. (\ref{eq:fourier_h}) is summed over the unit cells.
Integrating over the fast modes in the momentum shell and using momentum conservations in the free fermion model, the expression of the diagram in Fig. \ref{fig:renormalize1} is given by
\begin{flalign}
& \frac{1}{4} \sum_{\bar{m}=1}^4e^{i\frac{\pi}{2} \bar{m}i} \sum_{k q^\prime}  e^{i\vec{q}^\prime\cdot ia\hat{x} } h_{l}^\alpha(-q^\prime) c^\dagger (k+q^\prime+\frac{\pi}{2a}\bar{m}\hat{x}) \frac{1}{2}\sigma^\mu c(k)\nn\\
&\times e^{-i\vec{q}^\prime \cdot (j-l) a \hat{x}}
 \frac{1}{4N\beta}  \sum_{m=1}^4\sum_{k^\prime} e^{-i\frac{\pi}{2}(m+\bar{m})(j-l)} \nn\\
 &\times \langle c^\dagger(k^\prime-q^\prime)\frac{1}{2} \sigma^\nu c(k^\prime+\frac{\pi}{2a}(m+\bar{m})  \hat{x}) \nn\\
 &~~~~~~~~~~~~\cdot  c^\dagger(k^\prime+\frac{\pi }{2a} (m+\bar{m}) \hat{x})\frac{1}{2}\sigma^\alpha c(k^\prime-q^\prime)\rangle_{\text{f}},
\label{eq:renormalization_1_b_2}
\end{flalign}
in which $\vec{q}^\prime$ in $h_l^\alpha(-q^\prime)$ satisfies $|\vec{q}^\prime|\sim 0$, since $h_{l}^\alpha(n) $ is a smooth function of $n$.
In Eq. (\ref{eq:renormalization_1_b_2}), the factor $e^{-i\vec{q}^\prime \cdot (j-l) a \hat{x}}$ can be set as $1$ since it is a slowly varying variable.
Comparing with the following  Fourier representation of the magnetic field term in the action
\begin{flalign}
&\int d\tau \sum_n h_{i^\prime}^\alpha (\tau,n) S_{i^\prime+4n}^\alpha(\tau)=\nn\\
&\frac{1}{4}\sum_{kq}\sum_{m=1}^4 e^{i\vec{q}\cdot i^\prime a \hat{x} } e^{i\frac{\pi}{2}m i^\prime} h_{i^\prime}^\alpha(-q) c^\dagger(k+\frac{\pi}{2a}m\hat{x})\frac{1}{2}\sigma^\alpha c(k-q),
\label{eq:field_expr2_2}
\end{flalign}
it can be seen that Eq. (\ref{eq:renormalization_1_b_2}) is of the form in Eq. (\ref{eq:renormalize_diagram1}) which renormalizes $h_i^\mu$,
in which $\lambda_{jl}$ is given by Eq. (\ref{eq:lambda_coefficient2}).

\begin{figure}[h]
\begin{center}
\includegraphics[width=6cm]{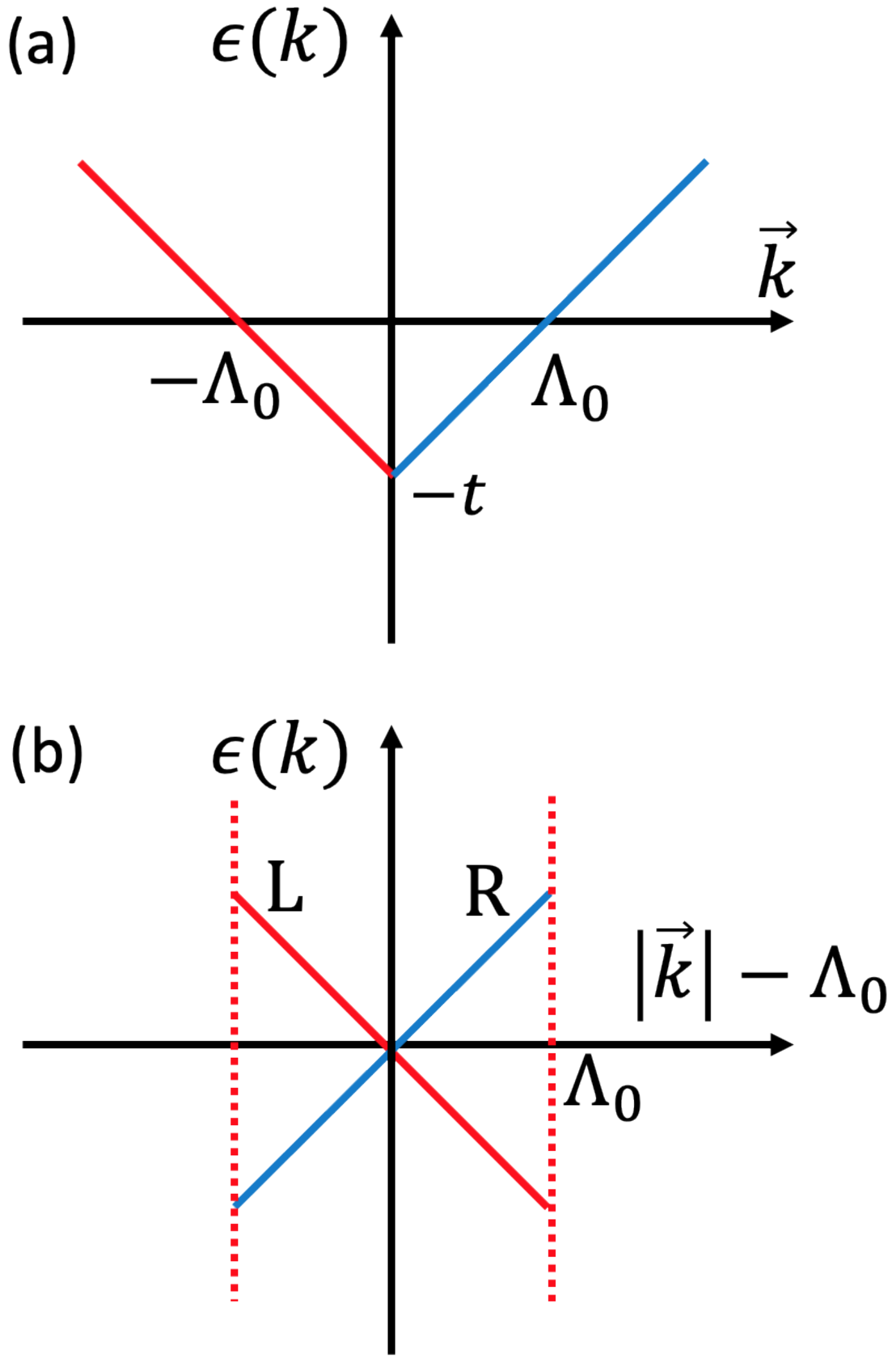}
\caption{(a) Linear dispersion of free fermion model,
(b) dispersion of the 1D Dirac fermion.
} \label{fig:dispersion}
\end{center}
\end{figure}

Eq. (\ref{eq:lambda_coefficient2}) is the desired expression for the coefficients $\lambda_{jl}$'s in the RG flow equations.
An analytic expression of $\lambda_{jl}$ is difficult, so we will  turn to numerical calculations. 
The numerical value of $\lambda_{jl}$ relies  on the underlying free fermion band structure $\epsilon(k)$, 
but the essential physics does not depend on the details of the band structure.
Hence,  the precise form of the  band structure is not essential  in our RG treatment.
The free fermion term $H_0$ in Eq. (\ref{eq:fermion_Ham}) has a $-t\cos(\vec{k}\cdot \hat{x})$ dispersion.
For simplicity, we modify the dispersion to a linear form, as
\bea
\epsilon(k)=v(|\vec{k}|-\Lambda_0),
\label{eq:linear_spectrum}
\eea
where $v=t/\Lambda_0$, and $\Lambda_0=\pi/(2a)$.
The figure for the spectrum in Eq. (\ref{eq:linear_spectrum}) is shown in Fig. \ref{fig:dispersion} (a).
The dispersion is essentially a Dirac fermion as shown in Fig. \ref{fig:dispersion} (b), in which the positions of the two gapless Fermi points are combined.

Next we evaluate the value of $\lambda_{jl}$ along RG flow.
Although the Dirac fermion has a cutoff $\Lambda_0$ in momentum space as shown in Fig. \ref{fig:dispersion} (b), the value of the Matsubara frequency at zero temperature is continuous and can extend to infinity.
Therefore, RG starts with an initial  cutoff $\Lambda_i\sim\infty$ in the frequency-momentum space,
and stops at $\Lambda_s\sim \Lambda_0/4$ as explained before.  
The values of $\lambda_{jl}$ in general depend on the cutoff $\Lambda=\Lambda_0/b$ where $\Lambda_0=\pi/(2a)$.
We note that $b$ can be smaller than $1$ 
since the value of the Matsubara frequency can take large values.


According to Fig. \ref{fig:dispersion} (b), the modes satisfying $\frac{\Lambda_0}{b+\Delta b}\leq \sqrt{(\omega/v)^2+(|\vec{k}|-\Lambda_0)^2}\leq \frac{\Lambda_0}{b}$ are integrated over.
The frequency and wavevector in the momentum shell can be parametrized as
\bea
\frac{\omega}{v\Lambda_0}=\frac{1}{b}\cos(\theta),~\frac{|\vec{k}|-\Lambda_0}{\Lambda_0}=\frac{1}{b}\sin(\theta),
\eea
where for each value of $|\vec{k}|-\Lambda_0\in[-\Lambda_0,\Lambda_0]$, we have both the left mover and the right mover.
We note that when $b<1$, $\theta$ cannot take all values in $[0,2\pi]$, since $-\Lambda_0\leq |\vec{k}|-\Lambda_0\leq \Lambda_0$.
On the other hand, when $b>1$, $\vec{k}$ cannot take all values in $[-\pi/a,\pi/a]$,  since some of the $\vec{k}$'s have been integrated over.

Then $\lambda_{jl}(b)$ as a function of $b$ can be obtained from Eq. (\ref{eq:lambda_coefficient2}) as
\begin{flalign}
&\lambda_{jl}(b)=-\frac{1}{64\pi t} \sum_m e^{-i\frac{\pi}{2}m(j-l)} \sum_{\nu=\pm 1} \int_0^{2\pi}d\theta f(\nu,m,\theta,b)\nn\\
&
\times \frac{1}{[i\cos\theta-b\bar{\epsilon}(b^{-1}\sin\theta+\nu)]}\nn\\
&\times \frac{1}{[i\cos\theta-b\bar{\epsilon}(b^{-1}\sin\theta+\nu+m)]},
\label{eq:lambda2_numerics}
\end{flalign}
in which  $\nu=1$ and $-1$ corresponds to the right and left movers in Fig. \ref{fig:dispersion} (b), respectively;
$f(\nu,m,\theta,b)$ is defined as
\bea
f(\nu,m,\theta,b)&=1,&\text{if } |b^{-1}\sin\theta|,|\bar{\epsilon}(b^{-1}\sin\theta+\nu)|,\nn\\
&&|\bar{\epsilon}(b^{-1}\sin\theta+\nu+m)| \leq\text{min}\{1,b^{-1}\},\nn\\ 
f(\nu,m,\theta,b)&=0,&~\text{otherwise},
\eea
imposing the condition that the magnitude of the spatial wavevector cannot exceed the cutoff;
and $\bar{\epsilon}$ is defined as
\bea
\bar{\epsilon}(x)=
|\text{mod}(x,4)|-1,
\eea
where $-2\leq\text{mod}(x,4)\leq 2$.

Next we write down the flow equations for $h_l^\alpha(b)$,
in which $b$ is the flow parameter, defined as $\Lambda(b)=\Lambda_0/b$ where $\Lambda(b)$ is the cutoff at the stage of the flow in consideration.
Since we are only interested in the $U(1)$ breaking effects, we neglect the renormalizations of the scaling fields due to the Hubbard term.
Although the Hubbard term also renormalizes the scaling fields, 
such renormalizations are SU(2) symmetric,
which does not affect the conclusions on U(1) breaking effects in the bosonization coefficients on a qualitative level.
We will also neglect the flows of the coupling constants $K+2J$ and $\Gamma$.
The reason is as follows.
As will be discussed in Sec. \ref{subsec:value_coeff},
the contributions to the bosonization coefficients from the $b\sim 0$ region (i.e., the $\Lambda(b) \sim \infty$ region) are negligible.
Hence it is enough to consider the RG flows within the range  $[b_i,b_s]$ where $b_i\sim O(1)$ and $b_s\sim 4$.
Since $K+2J$ and $\Gamma$ have scaling dimensions equal to zero and thereby are marginal operators, their flows can be safely neglected between the scales $b_i$ and $b_s$. 
On the other hand, in the high energy region $b\sim 0$ (i.e., $\Lambda(b)\sim \infty$),
there is no singularity in the perturbations, and as a result, $K(b_i)+2J(b_i)$ and $\Gamma(b_i)$ are analytic functions of the bare couplings  $K+2J$ and $J$.
Hence, in the weak coupling limit, it is enough to keep the leading order terms in $K(b_i)+2J(b_i)$ and $\Gamma(b_i)$, which are exactly given by $K+2J$ and $J$.
To summarize, according to the above arguments, $K(b)+2J(b)$ and $\Gamma(b)$ can be just taken as $K+2J$ and $J$ throughout the RG process in consideration. 

The flow equation of  $h_{l}^\mu$ ($1\leq l\leq 4$, $\mu=x,y,z$) up to one-loop level derived from the diagram in Fig. \ref{fig:renormalize1} is given by
\bea
\frac{dh_{l}^\mu}{d\ln b} &=& h_{l}^\mu -(K+2J)\sum_{\gamma,k}  \big[
\delta_{li} \delta_{\mu\gamma} \lambda_{j k} h_{k}^\gamma + \delta_{lj}\delta_{\mu\gamma}\lambda_{i k}h_{k}^\gamma
\big]\nn\\
&&-\Gamma\sum_{\gamma,k}\epsilon(\gamma)\big[
\delta_{li} \delta_{\mu\alpha}\lambda_{j k}h_{k}^\beta + \delta_{lj} \delta_{\mu\beta}\lambda_{i k}h_{k}^\alpha\nn\\
&&+\delta_{li} \delta_{\mu\beta}\lambda_{j k}h_{k}^\alpha + \delta_{lj} \delta_{\mu\alpha}\lambda_{i k}h_{k}^\beta
\big],
\label{eq:flow_h1_u}
\eea
in which the conventions are: $\gamma=x,y,\bar{x},\bar{y}$; $\alpha\neq \beta\neq \gamma$;
the spin direction index $\bar{x}$ (and $\bar{y}$) is identified with $x$ (and $y$) in the Kronecker delta and the scaling fields;
$\mathopen{<}ij\mathclose{>}=\gamma$; $i<j$; $1\leq i,j\leq 4$;
$5$ is identified with $1$.
The first term in Eq. (\ref{eq:flow_h1_u}) arises from the tree level scaling of the field $h_{l}^\mu$ 
(the dimension of the scaling field $h_{l}^\mu$ is $2-2[c^\dagger]=1$ where $[c^\dagger]=\frac{1}{2}$ is the dimension of the fermion operator at the free fermion fixed point),
whereas the second term is the one-loop correction.
Explicit expressions of the flow equations are included in Appendix \ref{app:explicit_RG}.

We note that Eq. (\ref{eq:flow_h1_u}) is invariant under the nonsymmorphic symmetry operations of the system, as proved in Appendix \ref{app:invariance}.


\subsection{Flow equations from the diagram in Fig. \ref{fig:renormalize2}}  
\label{sec:diagram_lambda3}

Next we consider the diagram in Fig. \ref{fig:renormalize2}.
This diagram gives rise to 
\bea
\lambda_{ilj} \Delta\ln b \int d\tau \sum_n h_l^\alpha(\tau_n) \frac{1}{2}c^\dagger_{i+4n} \sigma^\mu\sigma^\alpha\sigma^\nu c_{j+4n},
\eea
which leads to a renormalization of $h_{ij}^{\mu\cdot\alpha\cdot\nu}$, where the multiplication $\mu\cdot\nu$ ($\mu,\nu=1,2,3,4$) is defined as $x\cdot y=y\cdot x=z$, $x\cdot z=z\cdot x=y$, and $y\cdot z=z\cdot y=x$.
The coefficient $\lambda_{ilj}$ can be derived as 
\bea
\lambda_{ilj}\Delta \ln b&=&\frac{a}{8}\sum_m e^{i\frac{\pi}{2}m(l-i)}\int_{\Lambda/b}^{\Lambda}d^2k^\prime\nn\\
 &&\times e^{-i\vec{k}^\prime \cdot (i-j) a\hat{x}}\mathcal{G}(k^\prime) \mathcal{G}(k^\prime +\frac{\pi}{2a} m\hat{x}).
\label{eq:Lambda_3_expr2}
\eea
Details of the derivation of Eq. (\ref{eq:Lambda_3_expr2}) is included in Appendix \ref{app:diagram2}.

We demonstrate that the integration in  Eq. (\ref{eq:Lambda_3_expr2}) vanishes when $j=i\pm 1$, which applies to our case.
Notice that 
for $j=i\pm 1$, 
\bea
e^{-i(\vec{k}^\prime+\frac{\pi}{a}\hat{x}) \cdot (i-j) a\hat{x}}=-e^{-i\vec{k}^\prime \cdot (i-j) a\hat{x}}.
\eea
Then performing change of variables $(\omega\rightarrow -\omega,\vec{k}^\prime\rightarrow \vec{k}^\prime+\frac{\pi}{a}\hat{x})$ (the change of variable for $\omega$ is legitimate since $-\omega$ also lies in the momentum shell) and using $\epsilon(\vec{k}^\prime+\frac{\pi}{a}\hat{x})=-\epsilon(\vec{k}^\prime)$, 
it can be seen that the integration in Eq. (\ref{eq:Lambda_3_expr2}) changes sign, hence $\lambda_{ilj}=-\lambda_{ilj}=0$.
We note that when the Hamiltonian contains beyond nearest neighbor terms (e.g., $|j-i|=2$), the integration in Eq. (\ref{eq:Lambda_3_expr2}) no longer vanishes, and the diagram in Fig. \ref{fig:renormalize2} will contribute.

Because of the vanishing of $\lambda_{ilj}$,  the RG flow equations of $\lambda_{ilj}$ are
\bea
\frac{dh_{ij}^\mu}{d\ln b} = h_{ij}^\mu,
\label{eq:flow_eq_h2}
\eea
where $j= i \pm 1$.
Notice that initially $h_{ij}^{(0)\mu}=0$ at the beginning of the RG flow, hence the solution of Eq. (\ref{eq:flow_eq_h2}) is 
\bea
h_{i,i\pm 1}^\mu(b)= 0.
\label{eq:sol_RG_h2}
\eea

\subsection{Solving the RG flow equations}
\label{subsec:solve_RG}

The RG flow equations for $h_{ij}^\mu$ have already been solved in Eq. (\ref{eq:sol_RG_h2}).
To obtain $h^\alpha_j(b)$, the coupled RG flow equations in Eq. (\ref{eq:flow_h1_u}) need to be solved, which is a difficult problem.
Here we make the assumption that both $K+2J$ and $\Gamma$ are very small and only keep up to their first order terms.
With this  approximation, all the terms on the right hand side of the flow equations proportional to $K+2J$ or $\Gamma$ can be replaced with $bh^{(0)}$, where $h^{(0)}$ is the initial value (i.e., bare field) at the beginning of the RG flow $b_0$.
Here we note that $b_0$ in principle should be taken as $b_0=0$ since the Matsubara frequency can take infinite values. 

Within the first order approximation,  we obtain the following typical flow equation,
\bea
\frac{dh(x)}{dx}= h(x)+\lambda(x) e^x,
\label{eq:differential_eq}
\eea
where $x=\ln b$, and $\lambda$ is on order of $K+2J$ or $\Gamma$.
Let $h=ye^x$, Eq. (\ref{eq:differential_eq}) can be rewritten as
\bea
\frac{dy}{dx}=\lambda,
\eea
 which can be easily solved as 
 \bea
 y=y_0+\int dx \lambda(x).
 \eea
 Hence the solution of Eq. (\ref{eq:differential_eq}) is
 \bea
 h(b)=b[h^{(0)}+\int d \ln b\cdot  \lambda(b)].
 \label{eq:differential_eq_sol}
 \eea
 
Using Eq. (\ref{eq:differential_eq_sol}), Eq. (\ref{eq:flow_h1_u}) can be solved as
\begin{flalign}
&h_l^\mu(b)=b\big[h_{l}^{(0)\mu}\nn\\
& -(K+2J)\int d\ln b\sum_{\gamma,k}  (
\delta_{li} \delta_{\mu\gamma} \lambda_{j k} h_{k}^{(0)\gamma} + \delta_{lj}\delta_{\mu\gamma}\lambda_{i k}h_{k}^{(0)\gamma}
)\nn\\
&-\Gamma\int d\ln b\sum_{\gamma,k}\epsilon(\gamma)(
\delta_{li} \delta_{\mu\alpha}\lambda_{j k}h_{k}^{(0)\beta} + \delta_{lj} \delta_{\mu\beta}\lambda_{i k}h_{k}^{(0)\alpha}\nn\\
&+\delta_{li} \delta_{\mu\beta}\lambda_{j k}h_{k}^{(0)\alpha} + \delta_{lj} \delta_{\mu\alpha}\lambda_{i k}h_{k}^{(0)\beta}
)\big],
\label{eq:h_1_solved}
\end{flalign}
in which $\gamma=x,y,\bar{x},\bar{y}$; $\alpha\neq \beta\neq \gamma$;
the spin direction index $\bar{x}$ (and $\bar{y}$) is identified with $x$ (and $y$) in the Kronecker delta and the scaling field;
$\mathopen{<}ij\mathclose{>}=\gamma$; $i<j$; $1\leq i,j\leq 4$;
$5$ is identified with $1$;
and $h_{lm}^{(0)\mu}=0$ is used.



\section{Bosonization coefficients from RG flow equations}  
\label{sec:bosonize_coeff_from_RG}

In this section, we derive the nonsymmorphic bosonization coefficients from the  solutions of the RG flow equations.
Since $h^\mu_{i,i\pm 1}$'s are all zero as shown in  Eq. (\ref{eq:sol_RG_h2}), we will focus on the terms involving $h^\mu_{i}$.

\subsection{The uniform and staggered scaling fields at low energies}
\label{subsec:split_u_s}

Below the energy scale $\Lambda_s$, the differences among the sites within a unit cell are smeared out and our RG analysis stops.
At this stage, the coupling to the scaling fields is
\begin{flalign}
&-\sum_n\sum_{j=1}^4\sum_{\alpha=x,y,z}\int d\tau  h_j^\alpha (b;\tau,n) S_{j+4n}^\alpha(\tau),
\label{eq:coupling_Lambda_s}
\end{flalign}
in which $b>\Lambda_0/\Lambda_s$.
In addition, the $K+2J$ term  becomes  indistinguishable from the U(1) symmetric interaction $\frac{1}{2}(K+2J)\sum_i [S_i^xS_{i+1}^x+S_i^yS_{i+1}^y]$,
and the $\Gamma$ interaction cancels due to the $\epsilon(\gamma)$ factor in Eq. (\ref{eq:4rotated}).
Therefore, below the scale $\Lambda_s$, although the RG flow continues to renormalize the scaling fields,
such renormalizations respect the U(1) symmetry and there is no further U(1) breaking effect.
In view of this, for the purpose of a qualitative understanding of the U(1) breaking effects in the bosonization coefficients,
we will not discuss the flow equations below $\Lambda_s$,
bearing in mind that they only give rise to some overall U(1) preserving factors. 

When the cutoff is further lowered below $\Lambda_L$,
the only spin degrees of freedom are $S_u^\alpha$ and $S_s^\alpha$ defined in Eq. (\ref{eq:def_S_u_s}),
since the wavevectors far from $0$ and $\pi$ have all been integrated out.
We make a comment on the low energy field theory at the scale $\Lambda_L$.
As explained in Eq. (\ref{eq:energy_hierarchy}), below $\Lambda_L$, the fermion model can be approximated as a 1+1-dimensional Dirac fermion, and the spin-charge separation is applicable. 
The low energy field theory contains a spin part and a charge part.
The charge Hamiltonian has a $\cos(\sqrt{8\pi}\phi)$ term due to the repulsive Hubbard interaction where $\phi$ is the charge boson, which eventually opens a charge gap at the energy scale $m_c$ (where the mass acquires the same order of magnitude as the cutoff).
The spin Hamiltonian is of the XXZ type, since the smeared $K+2J$ interaction lowers the symmetry of the low energy Hamiltonian from SU(2) to U(1).
Clearly, the low energy theory has an emergent U(1) symmetry below the energy scale $\Lambda_L$.
It is worth to mention that although U(1) breaking renormalizations have already stopped at the scale $\Lambda_s$,
it is not legitimate to talk about a low energy theory at $\Lambda_s$, 
since $\Lambda_s\sim \Lambda_0/4$ is still in the high energy region.

To express Eq. (\ref{eq:coupling_Lambda_s}) in terms of $S_u^\alpha$ and $S_s^\alpha$ when the energy scale is below $\Lambda_L$, 
we should first project Eq. (\ref{eq:coupling_Lambda_s}) to left and right movers of the fermions,
and then rewrite the expression using $S_u^\alpha$ and $S_s^\alpha$.
Clearly, the projection of $S_{j+4n}^\alpha$  is given by
\bea
S_{j+4n}^\alpha(n)&=&\frac{1}{2}(c_L^\dagger \sigma^\alpha c_L+c_R^\dagger \sigma^\alpha c_R)\nn\\
&&+(-)^j \frac{1}{2}(c_L^\dagger \sigma^\alpha c_R+c_R^\dagger \sigma^\alpha c_L).
\label{eq:proj_LR_S1}
\eea
Plugging Eqs. (\ref{eq:S_us_c},\ref{eq:proj_LR_S1}) into Eq. (\ref{eq:coupling_Lambda_s}), we arrive at
\bea
-\sum_\alpha\int d\tau dx (h_u^\alpha S_u^\alpha+h_s^\alpha S_s^\alpha),
\label{eq:coupling_SuSs}
\eea
in which the uniform and staggered scaling fields $h_u^\alpha,h_s^\alpha$ are  given by
\bea
h_u^\alpha&=&\sum_{j=1}^4h_j^\alpha,\nn\\
h_s^\alpha&=&\sum_{j=1}^4 (-)^jh_j^\alpha.
\label{eq:expression_h_us}
\eea

\subsection{Derivations of the bosonization coefficients}

Plugging Eq. (\ref{eq:h_1_solved}) into Eq. (\ref{eq:expression_h_us}),
$h_u^\alpha$ and $h_s^\alpha$ (defined below the scale $\lambda_L$) can be expressed in terms of $h_j^{\mu(0)}$ as
\bea
h_u^\alpha&=&\sum_{\nu=x,y,z}\sum_{l=1}^4 \mathbb{D}_l^{\alpha\nu} h_l^{(0)\nu},\nn\\
h_s^\alpha&=&\sum_{\nu=x,y,z}\sum_{l=1}^4 (-)^l\mathbb{C}_l^{\alpha\nu} h_l^{(0)\nu},
\label{eq:relation_h_h0}
\eea
in which $\mathbb{D}_l^{\alpha\nu}$ and $\mathbb{C}_l^{\alpha\nu}$ are some numerical factors. 
Notice that the low energy fields $J^\alpha$ and $N^\alpha$ live at an energy scale below $m_c$ where the charge sector has been gapped out, leaving only the spin degrees of freedom. 
When the energy scale is further lowered from $\Lambda_L$ to below $m_c$,
$S_u^\alpha$ and $S_s^\alpha$ become just $J^\alpha$ and $N^\alpha$, respectively,
since $S_u^\alpha$ ($S_s^\alpha$) and $J^\alpha$ ($N^\alpha$) both correspond to the zero-  ($\pi$-) wavevector component of the low energy spin operator $\tilde{S}^\alpha_r(\tau)$ defined in Eq. (\ref{eq:def_S_u_s}).
As mentioned earlier, the RG flow between  $\Lambda_L$ and $m_c$ respects the emergent U(1) symmetry.
Hence, up to some additional U(1) symmetric renormalization factors,
the coupling to scaling fields in Eq. (\ref{eq:coupling_SuSs}) becomes
\bea
-\sum_\alpha \int d\tau dx (h_u^\alpha J^\alpha+h_s^\alpha N^\alpha)
\label{eq:coupling_JN}
\eea
below the scale $m_c$. 

Recall that performing functional derivatives $\partial/\partial h_j^\alpha$, $\partial/\partial h_u^\alpha$, and $\partial/\partial h_s^\alpha$ on the free energy can give the correlation functions involving $S_j^\alpha$, $J^\alpha$, and $N^\alpha$, respectively.
Using 
\bea
\frac{\partial }{\partial h_l^{(0)\nu}}=\sum_{\alpha=x,y,z}[\frac{\partial h_u^\alpha}{\partial h_l^{(0)\nu}}\frac{\partial }{\partial h_u^\alpha}+\frac{\partial h_s^\alpha}{\partial h_l^{(0)\nu}}\frac{\partial }{\partial h_s^\alpha}]
\label{eq:partial_derivative}
\eea
we see from Eq. (\ref{eq:relation_h_h0}) that
\bea
\mathbb{D}_l^{\alpha\nu}=\frac{\partial h_u^\alpha}{\partial h_l^{(0)\nu}},~\mathbb{C}_l^{\alpha\nu}=\frac{\partial h_s^\alpha}{\partial h_l^{(0)\nu}}.
\eea
In particular, it can be seen from Eq. (\ref{eq:partial_derivative}) that  $\mathbb{D}_l^{\alpha\nu}$ and $\mathbb{C}_l^{\alpha\nu}$ in Eq. (\ref{eq:relation_h_h0}) are related to the bosonization coefficients defined in Eq. (\ref{eq:nonsym_bosonization}) by
\bea
D_l^{\alpha\beta}=\mathbb{D}_l^{\beta\alpha},~
C_l^{\alpha\beta}=\mathbb{C}_l^{\beta\alpha}.
\label{eq:bosonize_DC_flower_DC}
\eea
$D_l^{\alpha\nu}$ and $C_l^{\alpha\nu}$ satisfy the symmetries in Eq. (\ref{eq:symmetries}),
since the RG flow equations are invariant under the symmetries.

Next we establish the precise relations between the bosonization coefficients and the solutions of the RG flow equations.
Notice that in Eq. (\ref{eq:h_1_solved}), $h_l^\mu(b)$ is linear in $h_j^{(0)\alpha}$.
Therefore, we have
\bea
h_i^\mu(b) &=& \sum_{\nu=x,y,z} \sum_{j=1}^4 E^{\mu\nu}_{ij} h_j^{(0)\nu},
\label{eq:EFG}
\eea
in which the explicit expressions of the coefficients $E^{\mu\nu}_{ij}$ can be read from Eq. (\ref{eq:h_1_solved}).
Plugging Eq. (\ref{eq:EFG}) into Eq. (\ref{eq:expression_h_us}) and comparing with Eqs. (\ref{eq:relation_h_h0},\ref{eq:bosonize_DC_flower_DC}), we obtain
\bea
D^{\alpha\beta}_l=\sum_{j=1}^4E^{\beta\alpha}_{jl},~
C^{\alpha\beta}_l=\sum_{j=1}^4 (-)^{l+j} E^{\beta\alpha}_{jl},
\label{eq:CD_EFG}
\eea
which give the bosonization coefficients via Eq. (\ref{eq:bosonize_DC_flower_DC}).

From  Eq. (\ref{eq:CD_EFG}),
the explicit expressions of the ten bosonization coefficients up to first orders in $K+2J$ and $\Gamma$ can be derived as
\bea
a_D&=&b [1-(K+2J)\int d\ln b \cdot
( \lambda_{11}+\lambda_{21} +\lambda_{31}+\lambda_{41})] \nn\\
b_D&=&0\nn\\
c_D&=& b \Gamma \int d \ln b\cdot(\lambda_{11}-\lambda_{21} -\lambda_{31}+\lambda_{41}) \nn\\
h_D&=&b\Gamma\int d \ln b\cdot (
\lambda_{11}-\lambda_{21}-\lambda_{31}+\lambda_{41}) \nn\\
i_D&=& b,
\label{eq:expression_D}
\eea
and
\bea
a_C&=&b[1-(K+2J)\int d \ln b \cdot
(\lambda_{11}-\lambda_{21}+\lambda_{31}-\lambda_{41})] \nn\\
b_C&=&0\nn\\
c_C&=& b\Gamma\int d\ln b\cdot (\lambda_{11}+\lambda_{21} -\lambda_{31}-\lambda_{41})\nn\\
h_C&=&b\Gamma\int d\ln b\cdot (\lambda_{11}+\lambda_{21} -\lambda_{31}-\lambda_{41}) \nn\\
i_C&=& b,
\label{eq:expression_C}
\eea
in which $\lambda_{ij}$'s are functions of $b$ as determined by Eq. (\ref{eq:lambda2_numerics}).

It can be seen from Eqs. (\ref{eq:expression_D},\ref{eq:expression_C}) that up to first order in $K+2J$ and $\Gamma$, the coefficients $b_D$ and $b_C$ vanish.
In fact, they start to appear at second order. 
Take $b_D$ as an example.
It can be observed from the flow equations that $h_j^y$ contributes to the flow of $h_j^z$, and $h_j^z$ contributes to the flow of $h_1^x$.
As a result, $h_1^x$ is affected by $h_j^y$, eventually leading to a nonzero $b_D$.
However, this is clearly a second order effect. 
Also notice that in Eqs. (\ref{eq:expression_D},\ref{eq:expression_C}), there are the relations $c_D=h_D$, $c_C=h_C$.
However, these equalities are not expected to hold when higher order terms are included.

We make some comments on the effects of the  RG flow below the energy scale $\Lambda_L$.
The scaling fields $h_\eta^\alpha(b)$ ($\eta=u,s$ and $\alpha=x,y,z$) are related to $h_\eta^\alpha(b_L)$ ($b_L=\Lambda_0/\Lambda_L\ll b$) via  the following relations
\bea
h_\eta^\alpha(b)&=&\sum_\beta M^{(\eta)}_{\alpha\beta}  h_\eta^\beta(b_L),
\eea
in which the matrix $M^{(\eta)}$ is a function of $b$ (for fixed $b_L$) and has U(1) symmetry since the U(1) breaking renormalization along the RG flow has already stopped at the scale $\Lambda_s$ (which is greater than $\Lambda_L$).
Using the chain rule of partial derivatives 
\bea
\frac{\partial }{\partial h_l^{(0)\nu}}=\sum_{\eta=u,s}\sum_{\alpha,\beta=x,y,z}\frac{\partial h_\eta^\beta(b_L)}{\partial h_l^{(0)\nu}}
\frac{\partial h_\eta^\alpha(b)}{\partial h_\eta^\beta(b_L)}
\frac{\partial }{\partial h_\eta^\alpha(b)}, 
\label{eq:chain_rule}
\eea
we see that the matrices $D_l(b)$, $C_l(b)$ are related to $D_l(b_L)$, $C_l(b_L)$ via
\bea
D_l(b) &=& D_l(b_L) (M^{(u)})^T,\nn\\
C_l(b) &=& C_l(b_L) (M^{(s)})^T.
\label{eq:C_D_b_bs}
\eea
When $b$ satisfies $\Lambda_0/b<m_c$, Eq. (\ref{eq:chain_rule}) produces the bosonization formulas,
and $D_l(b)$, $C_l(b)$ become the matrices of  bosonization coefficients in Eq. (\ref{eq:nonsym_bosonization}).
It is clear from Eq. (\ref{eq:C_D_b_bs}) that all the bosonization coefficients $D_l^{\alpha\beta}$, $C_l^{\alpha\beta}$ are affected by the RG flow below $\Lambda_L$, though in a U(1) invariant manner.
For example, in the special SU(2) case (i.e., the matrices $M^{(\eta)}$ have SU(2) symmetry, not just U(1) symmetry,
which applies to the SU(2)$_1$ line in Fig. \ref{fig:phase}),
$D_l^{\alpha\beta}=r^{(u)}D^{\alpha\beta}_l(b_L)$ and $C_l^{\alpha\beta}=r^{(s)}C^{\alpha\beta}_l(b_L)$ acquire an  overall renormalization  factor $r^{(u)}$ and $r^{(s)}$, respectively,
where $M^{(u)}_{\alpha\beta}=r^{(u)}\delta_{\alpha\beta}$ and $M^{(s)}_{\alpha\beta}=r^{(s)}\delta_{\alpha\beta}$.

\subsection{Values of the bosonization coefficients}
\label{subsec:value_coeff}

\begin{table}[!t]
\centering
 \begin{tabular}{| c || c | c | c | c | c | } 
 \hline
   & $|a_\Lambda|$ & $|i_\Lambda|$ & $|c_\Lambda|$ & $|h_\Lambda|$ & $|b_\Lambda|$ \\
 \hline
$\Lambda=C$ &  0.129 & 0.363 & 0.0244 & 0.0138 & 0.00103 \\
\hline
$\Lambda=D$ & 0.161 & 0.182 & 0.0359 & 0.0266 & ? \\
 \hline
\end{tabular}
\caption{Numerical values of $|w_\Lambda|$ ($w=a,i,c,h,b$; $\Lambda=C,D$) at the representative point $K+2J=1,J=-1,\Gamma=0.35$,
in which $|b_D|$ is too small and a reliable value cannot be extracted.
This table is taken from Ref. \onlinecite{Yang2022b}, where DMRG numerics are performed 
on a system of $L=144$ sites using periodic boundary conditions. 
}
\label{table:coeff}
\end{table}

From Eqs. (\ref{eq:expression_D},\ref{eq:expression_C}), it can be observed that $a_D,a_C \sim O(1)$ and $c_D,h_D,c_C,h_C\sim O(\Gamma)$,
whereas $b_D,b_C$ are second order in $K+2J$ and $\Gamma$.
Therefore, in the weak coupling limit (i.e., $|(K+2J)/J|,|\Gamma/J|\ll 1$), we have
\begin{flalign}
&|a_C|\sim|a_D|\gg |c_D|\sim|h_D|\sim|c_C|\sim|h_C|\gg |b_D|\sim|b_C|.
\label{eq:hierarchy}
\end{flalign}
Table \ref{table:coeff} is taken from Ref. \onlinecite{Yang2022b}, from which it can seen that the predicted hierarchy in Eq. (\ref{eq:hierarchy}) is indeed satisfied,
even though the value of  $K+2J$ is already large (equal to $1$).

\begin{figure}[h]
\begin{center}
\includegraphics[width=5.5cm]{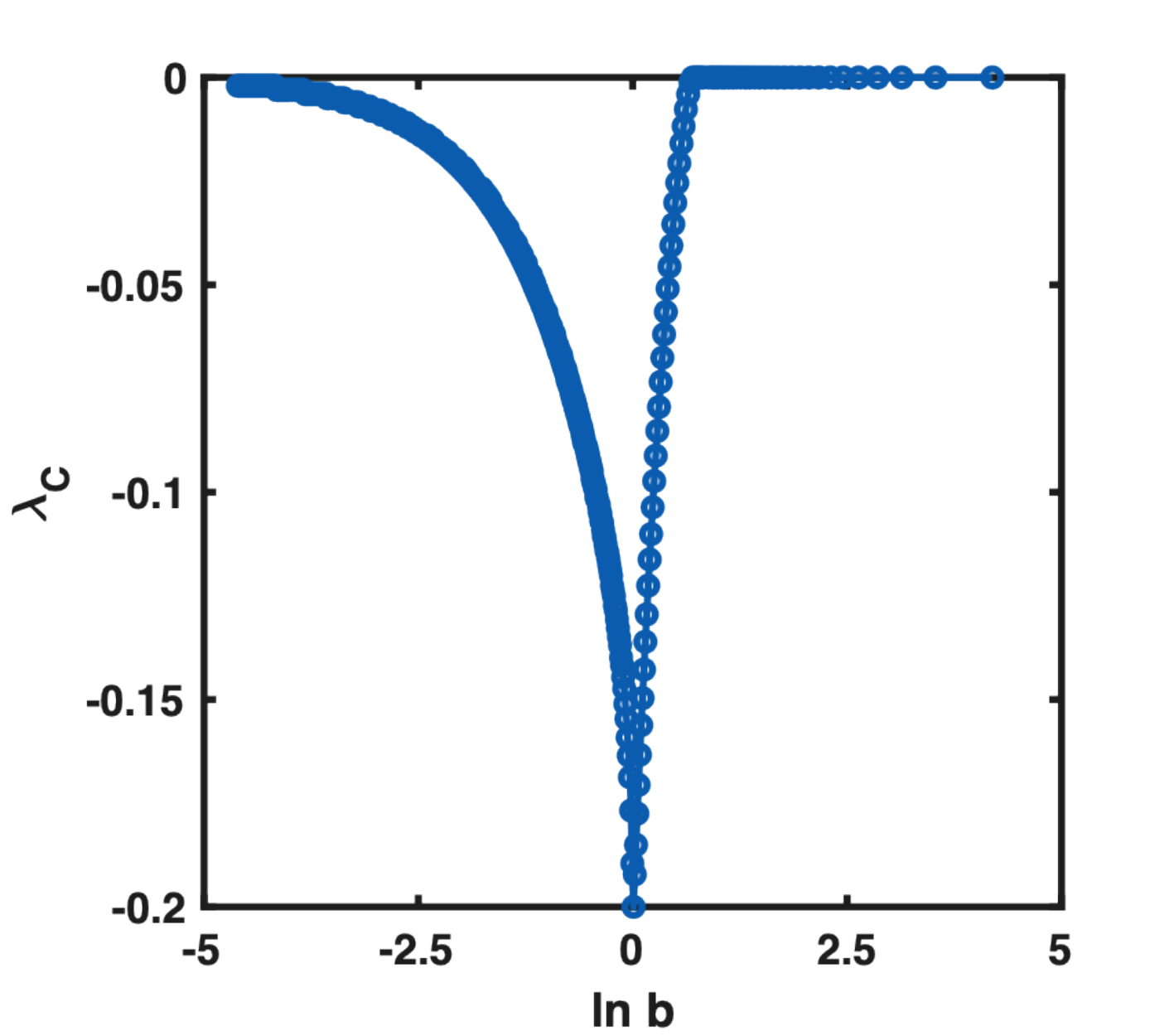}
\caption{$\lambda_C(b)$ as a function of $\ln b$, where the hopping $t$ is taken as $1$.
} \label{fig:lambda_C}
\end{center}
\end{figure}

Next we define $\lambda_D$ and $\lambda_C$ as
\bea
\lambda_D(b)&=&\lambda_{11}(b)-\lambda_{21}(b) -\lambda_{31}(b)+\lambda_{41}(b),\nn\\
\lambda_C(b)&=&\lambda_{11}(b)+\lambda_{21} (b)-\lambda_{31}(b)-\lambda_{41}(b).
\eea
Using $\lambda_{41}=\lambda_{45}$, as well as the inversion and translation symmetries, we obtain $\lambda_{21}=\lambda_{41}$,
which demonstrates that up to one-loop level there is the relation
\bea
\lambda_D(b)=\lambda_C(b).
\eea
Hence it enough to consider $\lambda_C(b)$.
Fig. \ref{fig:lambda_C} shows  $\lambda_C(b)$ as a function of $\ln b$ obtained by numerically calculating the integral in Eq. (\ref{eq:lambda2_numerics}),
and it can be seen that $\lambda_C(b)$ is always  negative.
We note that the integral  $\int d\ln b\cdot \lambda_C(b)$ converges
when $b$ is integrated from $0$ to $b_s\sim 4$.
When $b\ll 1$, the integration in Eq. (\ref{eq:lambda2_numerics}) is restricted within a narrow range $\theta\sim b$  due to the factor $f(\nu,m,\theta,b)$.
Let $x=\ln b$, and split $\int_{-\infty}^{x_s} dx \lambda_C(x)$  as $\int_{-\infty}^{y} dx \lambda_C(x)+\int_{y}^{x_s} dx \lambda_C(x)$ where $x_s\sim \ln 4$ and $y\ll 1$.
Since $\int_{-\infty}^{y} dx \lambda_C(x)$ goes like $\int_{-\infty}^{y} dx e^x$ which converges, we see that $\int_{-\infty}^{x_s} dx \lambda_C(x)$ is a converging integral. 
As a result, we conclude from Eq. (\ref{eq:expression_D}) and Eq. (\ref{eq:expression_C}) that RG predicts
\begin{flalign}
&c_C<0,~h_C<0,~c_D<0,~h_D<0.
\label{eq:predict_sign}
\end{flalign}

\subsection{Comparison with numerics}
\label{subsec:compare_numerics}

Next we check if the predictions in Eq. (\ref{eq:predict_sign}) are consistent with the numerical results. 
The method for numerically determining  the signs of the bosonization coefficients has been discussed in detail in Supplementary Materials in Ref. \onlinecite{Yang2022b}. 
In this subsection, we follow the method in Ref. \onlinecite{Yang2022b}.
We will focus on the ``C" coefficients,
since they correspond to $N^\alpha$ ($\alpha=x,y,z$) which are relevant operators and open a spin gap at low energies.
Appendix \ref{app:numerics_sign_D} discusses the numerical determinations of the ``D" coefficients, which are not successful, 
and the reasons remain not clear. 

Throughout this subsection, we work in the four-sublattice rotated frame and take the parameters as $K+2J=1$, $J=-1$, $\Gamma=0.35$ in accordance with Ref. \onlinecite{Yang2022b}.
DMRG numerical simulations are performed on a system of $L=144$ sites with periodic boundary conditions.
The bond dimension $m$ and truncation error $\epsilon$ in DMRG simulations are taken as $m=1400$ and $\epsilon=10^{-9}$.

Applying a small staggered magnetic field $h_\pi^z$ along $z$-direction, the low energy Hamiltonian can be derived as $-h_\pi^z i_C \int dxN^z$.
Since $N^z$ is a relevant operator, a spin gap opens and a nonzero expectation value $\langle N^z \rangle$ is developed in the low energy theory.
Using the nonsymmorphic bosonization formulas, the spin expectation values are 
\bea
\langle \vec{S}_{1+4n} \rangle &=& \langle N^z \rangle(-h_C,h_C,-i_C),\nn\\
\langle \vec{S}_{2+4n} \rangle &=& \langle N^z \rangle(-h_C,-h_C,i_C),\nn\\
\langle \vec{S}_{3+4n} \rangle &=& \langle N^z \rangle(h_C,-h_C,-i_C),\nn\\
\langle \vec{S}_{4+4n} \rangle &=& \langle N^z \rangle(h_C,h_C,i_C).
\label{eq:pattern_z_pi}
\eea
Taking $h_\pi^z=10^{-3}$, DMRG simulations are able to verify the pattern in Eq. (\ref{eq:pattern_z_pi}), with 
\bea
\langle N^z \rangle h_C&=&-4.5\times 10^{-4},\nn\\
\langle N^z \rangle i_C&=&0.0119.
\eea
This shows that 
\bea
h_C/i_C=-0.0378.
\label{eq:sign_hC}
\eea
Notice that $i_C$ is the dominant coefficient, and we expect that it does not change sign compared with the U(1) symmetric case for the microscopic Hamiltonian (i.e., when $K+2J=0$ and $\Gamma=0$).
Therefore, $h_C<0$ as determined from Eq. (\ref{eq:sign_hC}), which is consistent with the prediction in  Eq. (\ref{eq:predict_sign}).
In addition, Table \ref{table:coeff} gives a ratio $|h_C/i_C|$ equal to $0.0380$
where the values are obtained from studying spin correlation functions \cite{Yang2022b}.
It can be seen that the two approaches (magnetic field response vs. correlation functions) are fully consistent with each other. 

Next applying a small staggered magnetic field $h_\pi^x$ along $x$-direction, 
the low energy Hamiltonian can be derived as $-h_\pi^x (a_C \int dx N^x-b_D\int dx J^y)$.
Since the scaling dimension of $N^x$ is smaller than that of $J^y$, we expect that a nonzero expectation value $\langle N^x \rangle$ develops in the low energy theory.
Then the spin expectation values can be determined as follows from the nonsymmorphic bosonization formulas,
\bea
\langle \vec{S}_{1+4n} \rangle &=& \langle N^x \rangle(-a_C,-b_C,-c_C),\nn\\
\langle \vec{S}_{2+4n} \rangle &=& \langle N^x \rangle(a_C,-b_C,-c_C),\nn\\
\langle \vec{S}_{3+4n} \rangle &=& \langle N^x \rangle(-a_C,-b_C,c_C),\nn\\
\langle \vec{S}_{4+4n} \rangle &=& \langle N^x \rangle(a_C,-b_C,c_C).
\label{eq:pattern_hx_pi}
\eea
Since $a_C$ is the dominant coefficient, again $a_C$ is expected to be positive.
The patterns in Eq. (\ref{eq:pattern_hx_pi}) are verified by DMRG numerics, with the following values
\bea
\langle N^x \rangle a_C&=&0.0890,\nn\\
\langle N^x \rangle b_C&=&1.54\times 10^{-4},\nn\\
\langle N^x \rangle c_C&=&-0.00431,
\eea
which give the ratios as
\bea
b_C/a_C&=&0.0017,\nn\\
c_C/a_C&=&-0.0484.
\label{eq:sign_cC}
\eea
Hence the sign of $c_C$ is consistent with the prediction in  Eq. (\ref{eq:predict_sign}).
However, Table \ref{table:coeff} gives a ratio $|c_C/a_C|=0.189$, not consistent with the result in Eq. (\ref{eq:sign_cC}).
The reason for such discrepancy is unclear, and one possibility may be the neglection of the $J^y$ term in the analysis.

\section{Summary}
\label{sec:summary}

In summary, we have performed an RG study on the origin of the U(1) breaking terms in the bosonization formulas in the Luttinger liquid phase of the one-dimensional spin-1/2 Kitaev-Heisenberg-Gamma model with an antiferromagnetic Kitaev interaction. 
The RG analysis provides explanations for the origin of the ten non-universal bosonization coefficients in the abelian bosonization formulas of the spin operators.
It can also give predictions on the signs and order of magnitudes of these bosonization coefficients.
Our work is helpful to understand the rich physics related to nonsymmorphic symmetries in the gapless Luttinger liquid phases of the one-dimensional Kitaev spin models.

\begin{acknowledgments}
W.Y. and I.A. acknowledge support from NSERC Discovery Grant 04033-2016.
C.X. is partially supported by Strategic Priority Research Program of CAS (No. XDB28000000).
 A.N. acknowledges support from the Max Planck-UBC-UTokyo Center for Quantum Materials and the Canada First Research Excellence Fund
(CFREF) Quantum Materials and Future Technologies Program of the Stewart Blusson Quantum Matter Institute (SBQMI).
\end{acknowledgments}

\appendix

\begin{widetext}

\section{Explicit forms of the Hamiltonians}
\label{app:Ham}

The Hamiltonian in the unrotated frame is two-site periodic,  which has the form 
\begin{eqnarray}
H_{2n+1,2n+2}&=&K S_{2n+1}^x S_{2n+2}^x +\Gamma (S_{2n+1}^y S_{2n+2}^z+S_{2n+1}^z S_{2n+2}^y)+J\vec{S}_{2n+1}\cdot \vec{S}_{2n+2}, \nn\\
H_{2n+2,2n+3}&=&K S_{2n+2}^y S_{2n+3}^y +\Gamma (S_{2n+2}^z S_{2n+3}^x+S_{2n+2}^x S_{2n+3}^z)+J\vec{S}_{2n+2}\cdot \vec{S}_{2n+3}.
\end{eqnarray}
After the four-sublattice rotation, the Hamiltonian becomes four-site periodic, given by
\begin{eqnarray}
H^{\prime}_{4n+1,4n+2}&=& (K+2J)S_{4n+1}^xS_{4n+2}^x-J\vec{S}_{4n+1}\cdot\vec{S}_{4n+2}+\Gamma (S^y_{4n+1}S^z_{4n+2}+S^z_{4n+1}S^y_{4n+2}), \nn\\
H^{\prime}_{4n+2,4n+3}&=& (K+2J)S_{4n+2}^yS_{4n+3}^y-J\vec{S}_{4n+2}\cdot\vec{S}_{4n+3}+\Gamma (S^z_{4n+2}S^x_{4n+3}+S^x_{4n+2}S^z_{4n+3}), \nn\\
H^{\prime}_{4n+3,4n+4}&=& (K+2J)S_{4n+3}^xS_{4n+4}^x-J\vec{S}_{4n+3}\cdot\vec{S}_{4n+4}-\Gamma (S^y_{4n+3}S^z_{4n+4}+S^z_{4n+3}S^y_{4n+4}), \nn\\
H^{\prime}_{4n+4,4n+5}&=& (K+2J)S_{4n+4}^yS_{4n+5}^y-J\vec{S}_{4n+4}\cdot\vec{S}_{4n+5}-\Gamma (S^z_{4n+4}S^x_{4n+5}+S^x_{4n+4}S^z_{4n+5}).
\end{eqnarray}

\section{Explicit forms of the nonsymmorphic bosonization formulas}
\label{app:bosonization}

In the four-sublattice rotated frame, the explicit forms of the abelian bosonization formulas in the Luttinger liquid phase in Fig. \ref{fig:phase} are given by
\bea
S_{1+4n}^x&=&a_DJ^x+b_DJ^y+h_DJ^z-a_CN^x-b_CN^y-h_CN^z,\nn\\
S_{1+4n}^y&=&b_DJ^x+a_DJ^y-h_DJ^z-b_CN^x-a_CN^y+h_CN^z,\nn\\
S_{1+4n}^z&=&c_DJ^x-c_DJ^y+i_DJ^z-c_CN^x+c_CN^y-i_C N^z,
\label{eq:bosonization_S1}
\eea
\bea
S_{2+4n}^x&=&a_DJ^x-b_DJ^y-h_DJ^z+a_CN^x-b_CN^y-h_CN^z,\nn\\
S_{2+4n}^y&=&-b_DJ^x+a_DJ^y-h_DJ^z-b_CN^x+a_CN^y-h_CN^z,\nn\\
S_{2+4n}^z&=&-c_DJ^x-c_DJ^y+i_DJ^z-c_CN^x-c_CN^y+i_C N^z,
\label{eq:bosonization_S2}
\eea
\bea
S_{3+4n}^x&=&a_DJ^x+b_DJ^y-h_DJ^z-a_CN^x-b_CN^y+h_CN^z,\nn\\
S_{3+4n}^y&=&b_DJ^x+a_DJ^y+h_DJ^z-b_CN^x-a_CN^y-h_CN^z,\nn\\
S_{3+4n}^z&=&-c_DJ^x+c_DJ^y+i_DJ^z+c_CN^x-c_CN^y-i_C N^z,
\label{eq:bosonization_S3}
\eea
\bea
S_{4+4n}^x&=&a_DJ^x-b_DJ^y+h_DJ^z+a_CN^x-b_CN^y+h_CN^z,\nn\\
S_{4+4n}^y&=&-b_DJ^x+a_DJ^y+h_DJ^z-b_CN^x+a_CN^y+h_CN^z,\nn\\
S_{4+4n}^z&=&c_DJ^x+c_DJ^y+i_DJ^z+c_CN^x+c_CN^y+i_C N^z.
\label{eq:bosonization_S4}
\eea

\section{Nonabelian bosonization  of 1D repulsive Hubbard model at half filling}
\label{app:nonabelian_bosonization}

Here we give a quick review of the nonabelian bosonization method (for details, see Ref. \onlinecite{Affleck1988}).
The 1D spin-1/2 Dirac fermion exhibits the phenomenon of spin-charge separation and can be decomposed into an SU(2)$_1$ spin boson $g$ and a $U(1)$ charge boson $\phi$,
where the actions in real time for the SU(2) matrix $g$ and the real scalar $\phi$ are given by
\bea
S_g&=&\frac{1}{8\pi}\int d^2x \text{Tr}(\partial_\mu g^{-1}\partial^\mu g)+\frac{1}{12\pi}\int d^3 x\epsilon^{\mu\nu\lambda}\text{Tr}(\tilde{g}^{-1}\partial_\mu \tilde{g} \tilde{g}^{-1}\partial_\nu \tilde{g} \tilde{g}^{-1}\partial_\lambda \tilde{g}),\nn\\
S_\phi&=&\frac{1}{2}\int d^2x\partial_\mu\phi\partial^\mu\phi,
\eea
in which $\tilde{g}$ is an extension of $g$ from two-dimensional spacetime to three-dimension, 
and the velocities in $S_g$ and $S_\phi$ have been absorbed into a redefinition of time.
We note that because of the topological nature of the second term (i.e., WZW term) in $S_g$, the partition function does not depend on the way of extension.
In terms of $g$ and $\phi$, the hopping term between the  left and right movers can be bosonized as follows 
\bea
c_Lc_R^\dagger=\text{const.} ge^{i\sqrt{2\pi}\phi},
\label{eq:LR_bosonize}
\eea
where const. is a real constant.
When a repulsive Hubbard interaction $U>0$ is introduced, 
$S_g$ and $S_\phi$ are changed into
\begin{flalign}
&S_g^\prime=S_g+\frac{2U}{t}\int d^2x \vec{J}_L\cdot \vec{J}_R,\nn\\
&S_\phi^\prime=\frac{1}{2}(1+\frac{U}{2\pi t})\int d^2x\partial_\mu\phi\partial^\mu\phi-\lambda_\phi U\int d^2x \cos(\sqrt{8\pi}\phi),
\end{flalign}
in which $\lambda_\phi>0$ is a constant,
and the WZW current operators $\vec{J}_L$ and $\vec{J}_R$ are defined as
\bea
\vec{J}_L&=&\frac{i}{4\pi}\text{Tr}(\partial_+ g\cdot g^{-1}\vec{\sigma}),\nn\\
\vec{J}_R&=&-\frac{i}{4\pi}\text{Tr}(g^{-1}\partial_- g\vec{\sigma}),
\eea
where $\partial_{\pm}=\partial_t\pm\partial_x$.
It can be shown that the spin sector remains gapless since $\vec{J}_L\cdot \vec{J}_R$ is marginally irrelevant.
However, a gap opens in the charge sector since the $\cos(\sqrt{8\pi}\phi)$ term is relevant at low energies.
The scaling of the charge gap can be solved as $m_c\sim e^{-U/(\pi t)}$.

The above analysis shows that in the weak-$U$ limit, the low energy physics of the 1D repulsive Hubbard model is described by the SU(2)$_1$ WZW theory.
On the other hand, we know that according to the standard second order perturbation, the large-$U$ limit reduces to the SU(2) AFM Heisenberg model.
Since there is no phase transition between the weak-$U$ and large-$U$ limits,
the low energy physics of the SU(2) AFM Heisenberg model is also described by the SU(2)$_1$ WZW theory.
This provides a nonabelian bosonization description for the low energy physics of the AFM Heisenberg model in 1D. 

\section{Evaluation of Feynman diagrams}
\label{sec:diagram}

\subsection{Evaluation of diagram in Fig. \ref{fig:renormalize1}}
\label{app:diagram1}

We need to express the interactions and the spin operators in the frequency and momentum space.
The interaction term is 
\begin{flalign}
& \int d\tau \sum_n S_{i+4n}^\alpha(\tau) S_{j+4n}^\beta(\tau)=\nn\\
&\frac{1}{N\beta}\int d\tau \sum_n \sum_{k_1,k_2,k_3,k_4}  c^\dagger(k_1) \frac{1}{2}\sigma^\alpha c(k_2) \cdot c^\dagger(k_3) \frac{1}{2}\sigma^\beta c(k_4)
 e^{i(\omega_1-\omega_2+\omega_3-\omega_4) \tau}  e^{i(\vec{k}_1-\vec{k}_2)\cdot(i+4n)a\hat{x}} e^{i(\vec{k}_3-\vec{k}_4)\cdot(j+4n)a\hat{x}},
\label{eq:interaction_k}
\end{flalign}
in which $k=(i\omega,\vec{k})$ where $\omega$ is Matsubara frequency and $\vec{k}$ is the wavevector in space (we define the spatial wavevector as a vector to distinguish it from the spacetime combined index $k$, even though the system is 1D and $\vec{k}$ is in essence a scalar), $a$ is the lattice spacing,
$N$ is the system size, $\beta$ is the inverse of the temperature,
$n$ is summed over the unit cells, and $\hat{x}$ is the unit vector in the spatial direction.
Using the identity 
\bea
\frac{1}{N}\sum_n e^{i(\vec{k}_1-\vec{k}_2+\vec{k}_3-\vec{k}_4)\cdot 4na\hat{x}}=\frac{1}{4}\sum_{m=1}^4 \delta_{\vec{k}_1-\vec{k}_2+\vec{k}_3-\vec{k}_4,\frac{\pi}{2a}m\hat{x}},
\eea
Eq. (\ref{eq:interaction_k}) can be written as
\begin{flalign}
&\frac{\Delta}{N\beta}\frac{1}{4}\sum_m\sum_{k_1,k_2,k_3,k_4}  e^{i(\vec{k}_1-\vec{k}_2)\cdot(i-j) a\hat{x}}
e^{-i \frac{\pi }{2}m j} \delta_{k_1-k_2+k_3-k_4+\frac{\pi}{2a}m\hat{x},0}
 c^\dagger(k_1) \frac{1}{2}\sigma^\alpha c(k_2) \cdot c^\dagger(k_3) \frac{1}{2}\sigma^\beta c(k_4),
\end{flalign}
i.e.,
\bea
\frac{\Delta}{N\beta}\frac{1}{4}\sum_m\sum_{kk^\prime q} e^{i\vec{q}\cdot(i-j) a\hat{x}}e^{-i \frac{\pi }{2}m j}
c^\dagger(k+q) \frac{1}{2}\sigma^\alpha c(k) \cdot c^\dagger(k^\prime-q) \frac{1}{2}\sigma^\beta c(k^\prime+\frac{\pi}{2a}m\hat{x}),
\eea
where 
\bea
 c^\dagger(k)=\frac{1}{\sqrt{N\beta}}\int d\tau \sum_{j=1}^N  c^\dagger_j(\tau) e^{i(\omega\tau-\vec{k}\cdot ja\hat{x})}.
\eea
By defining the Fourier transform $h_l^\alpha(q)$ as
\bea
h_l^\alpha(q)=\frac{1}{\sqrt{N\beta}}\int d\tau \sum_{n=1}^{N/4} h_l^\alpha(\tau,n) e^{i(\omega\tau-\vec{q}\cdot4na\hat{x})}, 
\eea
the  coupling to the magnetic field becomes
\begin{flalign}
&\int d\tau \sum_n h_{l}^\alpha (\tau,n) S_{l+4n}^\alpha(\tau)
=\frac{1}{4}\sum_{kq}\sum_m e^{i\vec{q}\cdot la \hat{x} } e^{i\frac{\pi}{2}ml} h_l^\alpha(-q) c^\dagger(k+\frac{\pi}{2a}m\hat{x})\frac{1}{2}\sigma^\alpha c(k-q).
\label{eq:field_expr2}
\end{flalign}
Notice that $\vec{q}\in [0,\frac{\pi}{2a})$ in $h_l^\alpha(q)$ since $h_l^\alpha(n)$ is defined every four sites.


We set the momentum transfer $\vec{q}^\prime$ in $h_l^\alpha(q^\prime)$  as $|\vec{q}^\prime|\sim 0$ (both $S_u^\alpha$ and $S_s^\alpha$ correspond to $|\vec{q}^\prime|\sim 0$, which is the reason why they are not separated above the energy scale $\Lambda_s$).
The expression corresponding to the diagram in Fig. \ref{fig:renormalize1} is given by
\begin{flalign}
& \frac{1}{4N\beta} \sum_m e^{-i\frac{\pi}{2} mj} \sum_{kpq} e^{i\vec{q}\cdot(i-j)a\hat{x} } c^\dagger (k+q) \frac{1}{2}\sigma^\mu c(k) \cdot \frac{1}{4} \sum_{k^\prime q^\prime m^\prime} e^{i\vec{q}^\prime \cdot la\hat{x}} e^{i\frac{\pi}{2} m^\prime l} h_{l}^\alpha(-q^\prime)\nn\\
&\times
\langle c^\dagger(p-q) \frac{1}{2} \sigma^\nu c(p+\frac{\pi}{2a}m  \hat{x}) c^\dagger(k^\prime+\frac{\pi}{2a}  m^\prime\hat{x})\frac{1}{2}\sigma^\alpha c(k^\prime-q^\prime)\rangle_{\text{f}}.
\label{eq:expression_renorm1}
\end{flalign}
Since the  free fermion propagator is diagonal in  the frequency-momentum space, there are the following constraints   ($\bar{m}=1,2,3,4$)
\bea
m^\prime = m+\bar{m},~
p= k^\prime +\frac{\pi}{2a} \bar{m},~
q= q^\prime +\frac{\pi}{2a} \bar{m}.
\label{eq:momentum_conserv}
\eea 
Plugging Eq. (\ref{eq:momentum_conserv}) into Eq. (\ref{eq:expression_renorm1}) and rearranging the terms, we obtain the following alternative expression for Eq. (\ref{eq:expression_renorm1}), 
\begin{flalign}
& \frac{1}{4} \sum_{\bar{m}}e^{i\frac{\pi}{2} \bar{m}i} \sum_{k q^\prime}  e^{i\vec{q}^\prime\cdot ia\hat{x} } h_{l}^\alpha(-q^\prime) c^\dagger (k+q^\prime+\frac{\pi}{2a}\bar{m}\hat{x}) \frac{1}{2}\sigma^\mu c(k) e^{-i\vec{q}^\prime \cdot (j-l) a \hat{x}}\nn\\
&\times \frac{1}{4N\beta}  \sum_m\sum_{k^\prime} e^{-i\frac{\pi}{2}(m+\bar{m})(j-l)}
 \langle c^\dagger(k^\prime-q^\prime)\frac{1}{2} \sigma^\nu c(k^\prime+\frac{\pi}{2a}(m+\bar{m})  \hat{x}) c^\dagger(k^\prime+\frac{\pi }{2a} (m+\bar{m}) \hat{x})\frac{1}{2}\sigma^\alpha c(k^\prime-q^\prime)\rangle_{\text{f}}.
\label{eq:renormalization_1_b}
\end{flalign}
In Eq. (\ref{eq:renormalization_1_b}), the factor $e^{-i\vec{q}^\prime \cdot (j-l) a \hat{x}}$ can be set as $1$ since $\vec{q}^\prime$ is a slowly varying variable.
In fact, if we expand the exponential $e^{-i\vec{q}^\prime \cdot (j-l) a \hat{x}}$,  $|\vec{q}^\prime|^n$ becomes gradients in the real space, which renders the $n\neq 0$ terms less relevant than the leading $n=0$ term in the RG sense.
This justifies in a more rigorous way why $e^{-i\vec{q}^\prime \cdot (j-l) a \hat{x}}$ can be taken as $1$.
 
Then by using Eq. (\ref{eq:field_expr2}), it can be checked that Eq. (\ref{eq:renormalization_1_b}) becomes
\bea
\lambda_{jl}\delta_{\nu\alpha}\Delta \ln b\cdot \int d\tau \sum_n h_{l}^\mu (\tau,n) S_{u,l}^\mu(\tau,n),
\eea
in which 
\begin{flalign}
\lambda_{jl} \Delta\ln b =  -\frac{a}{8}  \sum_m e^{-i\frac{\pi}{2}m(j-l)}\int_{\Lambda/b^\prime}^{\Lambda/b} d^2k  \mathcal{G}(k) \mathcal{G}(k+\frac{\pi }{2a} m \hat{x}),
\label{eq:lambda_coefficient}
\end{flalign}
where $\mathcal{G}(k)$ is the free fermion Green's function defined as
\bea
\mathcal{G}(k)=\frac{1}{i\omega-\epsilon(k)}. 
\label{eq:free_f_green}
\eea
In Eq. (\ref{eq:free_f_green}), $\epsilon(k)$ is the free fermion dispersion which includes the chemical potential term. 

\subsection{Evaluation of diagram in Fig. \ref{fig:renormalize2}}
\label{app:diagram2}

The expression corresponding to Fig. \ref{fig:renormalize2} is
\bea
&\frac{1}{16N\beta} \sum_{kpq}\sum_{m}\sum_{k^\prime q^\prime}\sum_{m^\prime} e^{-i\frac{\pi}{2}mj} e^{i\frac{\pi}{2}m^\prime l} 
e^{i\vec{q}\cdot (i-j)a\hat{x}}  e^{i\vec{q}^\prime \cdot la\hat{x}} h_l^\alpha (-q^\prime)\nn\\
&\times c^\dagger(k+q) \frac{1}{2} \sigma^\mu \langle c(k) c^\dagger (k^\prime +\frac{\pi}{2a}m^\prime) \frac{1}{2} \sigma^\alpha c(k^\prime-q^\prime)c^\dagger(p-q-\frac{\pi}{2a}m)\rangle_f \frac{1}{2} \sigma^\nu c(p).
\label{eq:renorm2_a}
\eea
Momentum conservation requires 
\bea
k&=& k^\prime +\frac{\pi}{2a} (m+\bar{m}),\nn\\
q&=& p-k^\prime+q^\prime-\frac{\pi}{2a}m,\nn\\
m^\prime&=&m+\bar{m}.
\eea
Then it can be shown that Eq. (\ref{eq:renorm2_a}) becomes
\bea
&\frac{1}{16N\beta} \sum_{pq^\prime}\sum_{\bar{m}} 
e^{i\frac{\pi}{2}\bar{m}i} e^{i\vec{p}\cdot (i-j)a\hat{x}} e^{i\vec{q}^\prime \cdot (i-j+l)a\hat{x}} h_l^\alpha (-q^\prime) 
c^\dagger(p+q^\prime+\frac{\pi}{2a}\bar{m}) \frac{1}{2} \sigma^\mu\sigma^\alpha\sigma^\nu c(p)\nn\\
&\times \frac{1}{4} \sum_{k^\prime} \sum_{m} e^{-i\vec{k}^\prime \cdot (i-j) a\hat{x}} e^{i\frac{\pi}{2}(m+\bar{m}) (l-i)}
2\mathcal{G}(k^\prime +\frac{\pi}{2a} (m+\bar{m})\hat{x}) \mathcal{G}(k^\prime-q^\prime),
\label{eq:renorm2_b}
\eea
in which the factor of two before the Green's function comes from the sum over the spin degree of freedom, and $\sum_{k^\prime}$ is restricted within the momentum shell, i.e., the fast modes.
Notice that 
\begin{flalign}
\int d\tau \sum_n c^\dagger_{i+4n} \frac{1}{2}\sigma^\rho c_{j+4n} h_{ij}^\rho (n) 
=\frac{1}{4N\beta} \sum_{pq^\prime} \sum_{\bar{m}} e^{i\vec{p}\cdot (i-j)a\hat{x}} e^{i\vec{q}^\prime\cdot  ia\hat{x}}e^{i\frac{\pi}{2}\bar{m}i} h_{ij} (-q^\prime) c^\dagger(p+q^\prime+\frac{\pi}{2a}\bar{m}) \frac{1}{2}\sigma^\rho c(p).
\label{eq:renorm2_c}
\end{flalign}
Plugging Eq. (\ref{eq:renorm2_c}) into Eq. (\ref{eq:renorm2_b}) and neglecting the $e^{i\vec{q}^\prime\cdot  (-j+l)a\hat{x}}$ factor in Eq. (\ref{eq:renorm2_b}) since $\vec{q}^\prime$ is a very small wavevector, 
we obtain
\bea
\lambda_{ilj} \Delta\ln b \int d\tau \sum_n h_{ij}^{\mu\cdot\alpha\cdot\nu} S_{l+4n}^{\mu\cdot\alpha\cdot\nu} (\tau),
\eea
where the coefficient $\lambda_{ilj}$ is
\bea
\lambda_{ilj}\Delta \ln b=\frac{a}{8}\sum_m e^{i\frac{\pi}{2}m(l-i)}\int_{\Lambda/b}^{\Lambda}d^2k^\prime
e^{-i\vec{k}^\prime \cdot (i-j) a\hat{x}} \mathcal{G}(k^\prime) \mathcal{G}(k^\prime +\frac{\pi}{2a} m\hat{x}).
\label{eq:Lambda_3_expr}
\eea
Notice that shifting $l$ by a multiple of $4$ does not affect the result in Eq. (\ref{eq:Lambda_3_expr}),
hence we can impose the condition $l\geq \min\{i,j\}$.
Apparently, Eq. (\ref{eq:Lambda_3_expr}) is invariant under spatial translation ($i,j,l\rightarrow i+t,j+t,l+t$) and inversion ($i,j,l\rightarrow -i,-l,-j$), as it must be.


\section{Explicit RG flow equations}
\label{app:explicit_RG}

The explicit flow equations for $h_l^\mu$ ($1\leq l\leq 4$ and $\mu=x,y,z$) are
\bea
\frac{dh_1^x}{d\ln b}&=& h_1^x-(K+2J) \sum_j \lambda_{2j}h_j^x+\Gamma\sum_j \lambda_{4j}h_j^z,\nn\\
\frac{dh_1^y}{d\ln b}&=& h_1^y-(K+2J) \sum_j \lambda_{4j}h_j^y-\Gamma\sum_j \lambda_{2j}h_j^z,\nn\\
\frac{dh_1^z}{d\ln b}&=& h_1^z+\Gamma \sum_j \lambda_{4j}h_j^x-\Gamma \sum_j \lambda_{2j}h_j^y,
\eea
\bea
\frac{dh_2^x}{d\ln b}&=& h_2^x-(K+2J) \sum_j \lambda_{1j}h_j^x-\Gamma\sum_j \lambda_{3j}h_j^z,\nn\\
\frac{dh_2^y}{d\ln b}&=& h_2^y-(K+2J) \sum_j \lambda_{3j}h_j^y-\Gamma\sum_j \lambda_{1j}h_j^z,\nn\\
\frac{dh_2^z}{d\ln b}&=& h_2^z-\Gamma \sum_j \lambda_{3j}h_j^x-\Gamma \sum_j \lambda_{1j}h_j^y,
\eea
\bea
\frac{dh_3^x}{d\ln b}&=& h_3^x-(K+2J) \sum_j \lambda_{4j}h_j^x-\Gamma\sum_j \lambda_{2j}h_j^z,\nn\\
\frac{dh_3^y}{d\ln b}&=& h_3^y-(K+2J) \sum_j \lambda_{2j}h_j^y+\Gamma\sum_j \lambda_{4j}h_j^z,\nn\\
\frac{dh_3^z}{d\ln b}&=& h_3^z-\Gamma \sum_j \lambda_{2j}h_j^x+\Gamma \sum_j \lambda_{4j}h_j^y,
\eea
\bea
\frac{dh_4^x}{d\ln b}&=& h_4^x-(K+2J) \sum_j \lambda_{3j}h_j^x+\Gamma\sum_j \lambda_{1j}h_j^z,\nn\\
\frac{dh_4^y}{d\ln b}&=& h_4^y-(K+2J) \sum_j \lambda_{1j}h_j^y+\Gamma\sum_j \lambda_{3j}h_j^z,\nn\\
\frac{dh_4^z}{d\ln b}&=& h_4^z+\Gamma \sum_j \lambda_{1j}h_j^x+\Gamma \sum_j \lambda_{3j}h_j^y,
\eea
in which $\lambda_{41}$ and $\lambda_{14}$ should be understood as $\lambda_{45}$ and $\lambda_{54}$, respectively.

\section{Invariance of RG flow equations under symmetries}  
\label{app:invariance}
  
Apparently,  Eqs. (\ref{eq:flow_h1_u}) does not respect the U(1) symmetry. 
We will verify that they are invariant under the nonsymmorphic symmetries in Eq. (\ref{eq:symmetries}).

1) Eq. (\ref{eq:flow_h1_u}) remains invariant under $h_{l}^\mu\rightarrow -h_{l}^\mu$,
hence time reversal symmetry is satisfied. 

2) Suppose we shift $l$ to $l+1$, and perform the rotation on spin indices by $R_z=R(\hat{z},-\frac{\pi}{2})$, 
then Eq. (\ref{eq:flow_h1_u}) becomes
\bea
\frac{dh_{l+1}^{R_z\mu}}{d\ln b} &=& h_{l+1}^{R_z\mu} -\sum_{\gamma,k} (K+2J) \big[
\delta_{l+1,i} \delta_{R_z\mu,\gamma} \lambda_{j k} h_{k}^{\gamma} + \delta_{l+1,j}\delta_{R_z\mu,\gamma}\lambda_{i k}h_{k}^{\gamma}
\big]\nn\\
&&-\sum_{\gamma,k}\epsilon(\gamma)\big[
\delta_{l+1,i} \delta_{R_z\mu,\alpha}\lambda_{j k}h_{k}^{\beta} + \delta_{l+1,j} \delta_{R_z\mu,\beta}\lambda_{i k}h_{k}^{\alpha}
+\delta_{l+1,i} \delta_{R_z\mu,\beta}\lambda_{j k}h_{k}^{\alpha} + \delta_{l+1,j} \delta_{R_z\mu,\alpha}\lambda_{i k}h_{k}^{ \beta}
\big].\nn\\
\label{eq:flow_h1_u_Rz}
\eea
Define $R_z^\prime$ as $R_z^\prime(x,y,\bar{x},\bar{y})=(y,\bar{x},\bar{y},x)$.
Then $\mathopen{<}i+1,j+1\mathclose{>}=R^\prime_z\gamma$.
Notice that $R_z\alpha\neq R_z\beta\neq R_z\gamma$ if $\alpha\neq \beta\neq \gamma$, and $\epsilon(R_z^\prime\gamma)=(-)^{i-1}\epsilon(\gamma)$ where $\gamma=<i,j>$.
Then we can change the dummy variable from $\gamma$ to $R_z^\prime\gamma$ and also from $k$ to $k+1$. 
As a result, Eq. (\ref{eq:flow_h1_u_Rz}) becomes
\bea
\frac{dh_{l+1}^{R_z\mu}}{d\ln b} &=& h_{l+1}^{R_z\mu} -\sum_{\gamma,k} (K+2J) \big[
\delta_{l+1,i+1} \delta_{R_z\mu,R_z\gamma} \lambda_{j+1, k+1} h_{k+1}^{R_z\gamma} + \delta_{l+1,j+1}\delta_{R_z\mu,R_z\gamma}\lambda_{i+1, k+1}h_{k+1}^{R_z\gamma}
\big]\nn\\
&&-(-)^{i-1}\sum_{\gamma,k}\epsilon(\gamma)\big[
\delta_{l+1,i+1} \delta_{R_z\mu,R_z\alpha}\lambda_{j+1, k+1}h_{k+1}^{R_z\beta} + \delta_{l+1,j+1} \delta_{R_z\mu,R_z\beta}\lambda_{i+1, k+1}h_{k+1}^{R_z\alpha}\nn\\
&&
+\delta_{l+1,i+1} \delta_{R_z\mu,R_z\beta}\lambda_{j+1, k+1}h_{k+1}^{R_z\alpha} + \delta_{l+1,j+1} \delta_{R_z\mu,R_z\alpha}\lambda_{i+1, k+1}h_{k+1}^{R_z \beta}
\big],
\label{eq:flow_h1_u_Rz_2}
\eea
in which $i$, $j$, $\alpha$, $\beta$ have been changed to $i+1$, $j+1$, $R_z\alpha$, $R_z\beta$, respectively, in accordance with changing $\gamma$ to $R_z\gamma$.
Using  $\lambda_{jk}=\lambda_{j+1,k+1}$ and $\delta_{\mu\gamma}=\delta_{R_z\mu,R_z\gamma}$, we obtain
\bea
\frac{dh_{l+1}^{R_z\mu}}{d\ln b} &=& h_{l+1}^{R_z\mu} -\sum_{\gamma,k} (K+2J) \big[
\delta_{li} \delta_{\mu\gamma} \lambda_{jk} h_{k+1}^{R_z\gamma} + \delta_{lj}\delta_{\mu\gamma}\lambda_{i k}h_{k+1}^{R_z\gamma}
\big]\nn\\
&&-\sum_{\gamma,k}(-)^{i-1}\epsilon(\gamma)\big[
\delta_{li} \delta_{\mu\alpha}\lambda_{jk}h_{k+1}^{R_z\beta} + \delta_{lj} \delta_{\mu\beta}\lambda_{ik}h_{k+1}^{R_z\alpha}
+\delta_{li} \delta_{\mu\beta}\lambda_{jk}h_{k+1}^{R_z\alpha} + \delta_{lj} \delta_{\mu\alpha}\lambda_{i k}h_{k+1}^{R_z \beta}
\big].\nn\\
\label{eq:flow_h1_u_Rz_3}
\eea

Using the fact that the site index $i$ in Eq. (\ref{eq:flow_h1_u_Rz_3}) is the left point of the bond $\gamma$, we see that 
\bea
(-)^{i-1}=(-)^{\delta_{\alpha,x}+\delta_{\beta,x}}.
\eea
Hence, we note that the sign factor $(-)^{i-1}$ in the second term in the right hand side of Eq. (\ref{eq:flow_h1_u_Rz_3}) can be replaced by either $(-)^{\delta_{\mu,x}+\delta_{\beta,x}}$ (for the $h_{k+1}^{R_z\beta}$ term), or $(-)^{\delta_{\mu,x}+\delta_{\alpha,x}}$ (for the $h_{k+1}^{R_z\alpha}$ term),
since in whichever case the sign factor is $(-)^{\delta_{\alpha,x}+\delta_{\beta,x}}$ because of the Kronecker delta $\delta_{\mu\alpha}$ (for the $h_{k+1}^{R_z\beta}$ term), and $\delta_{\mu\beta}$ (for the  $h_{k+1}^{R_z\alpha}$ term).
Therefore, Eq. (\ref{eq:flow_h1_u_Rz_3}) can be re-written as
\bea
\frac{dh_{l+1}^{R_z\mu}}{d\ln b} &=& h_{l+1}^{R_z\mu} -\sum_{\gamma,k} (K+2J) \big[
\delta_{li} \delta_{\mu\gamma} \lambda_{jk} h_{k+1}^{R_z\gamma} + \delta_{lj}\delta_{\mu\gamma}\lambda_{i k}h_{k+1}^{R_z\gamma}
\big]\nn\\
&&-\sum_{\gamma,k}\epsilon(\gamma)\big[
\delta_{li} \delta_{\mu\alpha}\lambda_{jk}(-)^{\delta_{\mu,x}+\delta_{\beta,x}}h_{k+1}^{R_z\beta} + \delta_{lj} \delta_{\mu\beta}\lambda_{ik}(-)^{\delta_{\mu,x}+\delta_{\alpha,x}}h_{k+1}^{R_z\alpha}\nn\\
&&+\delta_{li} \delta_{\mu\beta}\lambda_{jk}(-)^{\delta_{\mu,x}+\delta_{\alpha,x}}h_{k+1}^{R_z\alpha} + \delta_{lj} \delta_{\mu\alpha}\lambda_{i k}(-)^{\delta_{\mu,x}+\delta_{\beta,x}}h_{k+1}^{R_z \beta}
\big].
\label{eq:flow_h1_u_Rz_3}
\eea

On the other hand, using $R_z(S^x,S^y,S^z)\rightarrow (-S^y,S^x,S^z)$,
it can be seen that the invariance of the RG flow equations under the symmetry operation $R_zT_a$ exactly requires Eq. (\ref{eq:flow_h1_u_Rz_3}).
Hence we conclude that the flow equations have the symmetry imposed by $R_zT_a$.
 
3) Suppose we perform an operation on $h_i^\alpha$ ($1\leq i\leq 4$, $\alpha=x,y,z$) as
\bea
h_i^\alpha\rightarrow h_{5-i}^{R_y\alpha},
\eea 
in which $R_y$ is $R(\hat{y},\pi)$ for short.
Then Eq. (\ref{eq:flow_h1_u}) becomes
\bea
\frac{dh_{5-l}^{R_y\mu}}{d\ln b} &=& h_{5-l}^{R_y\mu} -\sum_{\gamma,k} (K+2J) \big[
\delta_{5-l,i} \delta_{R_y\mu,\gamma} \lambda_{j k} h_{k}^{R_y\gamma} + \delta_{5-l,j}\delta_{R_y\mu,\gamma}\lambda_{i k}h_{k}^{R_y\gamma}
\big]\nn\\
&&-\sum_{\gamma,k}\epsilon(\gamma)\big[
\delta_{5-l,i} \delta_{R_y\mu,\alpha}\lambda_{j k}h_{k}^{R_y\beta} + \delta_{5-l,j} \delta_{R_y\mu,\beta}\lambda_{i k}h_{k}^{R_y\alpha}
\nn\\
&&+\delta_{5-l,i} \delta_{R_y\mu,\beta}\lambda_{j k}h_{k}^{R_y \alpha} + \delta_{5-l,j} \delta_{R_y\mu,\alpha}\lambda_{i k}h_{k}^{R_y \beta}
\big].
\label{eq:flow_h1_u_Ry}
\eea
Define $R_y^\prime$ as $R_y^\prime(x,y,\bar{x},\bar{y})=(\bar{x},y,x,\bar{y})$.
Then $\mathopen{<}5-i,5-j\mathclose{>}=R^\prime_y\gamma$.
Notice that $R_y\alpha\neq R_y\beta\neq R_y\gamma$ if $\alpha\neq \beta\neq \gamma$, and $\epsilon(R_y^\prime\gamma)=(-)^{i}\epsilon(\gamma)$ where $\gamma=<i,j>$.
Then we can change the dummy variable from $\gamma$ to $R_y^\prime\gamma$ and also from $k$ to $5-k$. 
As a result, Eq. (\ref{eq:flow_h1_u_Ry}) becomes
\bea
\frac{dh_{5-l}^{R_y\mu}}{d\ln b} &=& h_{5-l}^{R_y\mu} -\sum_{\gamma,k} (K+2J) \big[
\delta_{5-l,5-i} \delta_{R_y\mu,R_y\gamma} \lambda_{5-j,5-k} h_{5-k}^{R_y\gamma} + \delta_{5-l,5-j}\delta_{R_y\mu,R_y\gamma}\lambda_{5-i,5-k}h_{5-k}^{R_y\gamma}
\big]\nn\\
&&-\sum_{\gamma,k}(-)^i\epsilon(\gamma)\big[
\delta_{5-l,5-i} \delta_{R_y\mu,R_\alpha}\lambda_{5-j,5-k}h_{5-k}^{R_y\beta} + \delta_{5-l,5-j} \delta_{R_y\mu,R_y\beta}\lambda_{5-i,5-k}h_{5-k}^{R_y\alpha}\nn\\
&&+\delta_{5-l,5-i} \delta_{R_y\mu,R_y\beta}\lambda_{5-j,5-k}h_{5-k}^{R_y \alpha} + \delta_{5-l,5-j} \delta_{R_y\mu,R_y\alpha}\lambda_{5-i,5-k}h_{5-k}^{R_y \beta}
\big],
\label{eq:flow_h1_u_Ry_2}
\eea
in which $i$, $j$, $\alpha$, $\beta$ have been changed to $5-j$, $5-i$, $R_y\alpha$, $R_y\beta$, respectively, in accordance with changing $\gamma$ to $R_z\gamma$.
Using  $\lambda_{jk}=\lambda_{5-j,5-k}$ and $\delta_{\mu\gamma}=\delta_{R_y\mu,R_y\gamma}$, we obtain
\bea
\frac{dh_{5-l}^{R_y\mu}}{d\ln b} &=& h_{5-l}^{R_y\mu} -\sum_{\gamma,k} (K+2J) \big[
\delta_{li} \delta_{\mu\gamma} \lambda_{jk} h_{5-k}^{R_y\gamma} + \delta_{lj}\delta_{\mu\gamma}\lambda_{ik}h_{5-k}^{R_y\gamma}
\big]\nn\\
&&-\sum_{\gamma,k}(-)^i\epsilon(\gamma)\big[
\delta_{li} \delta_{\mu\alpha}\lambda_{jk}h_{5-k}^{R_y\beta} + \delta_{lj} \delta_{\mu\beta}\lambda_{ik}h_{5-k}^{R_y\alpha}+\delta_{li} \delta_{\mu\beta}\lambda_{jk}h_{5-k}^{R_y \alpha} + \delta_{lj} \delta_{\mu\alpha}\lambda_{ik}h_{5-k}^{R_y \beta}
\big].
\label{eq:flow_h1_u_Ry_3}
\eea

Using the fact that the site index $i$ in Eq. (\ref{eq:flow_h1_u_Rz_3}) is the left point of the bond $\gamma$, we see that 
\bea
(-)^{i}=(-)^{\delta_{\alpha,x}+\delta_{\alpha,z}+\delta_{\beta,x}+\delta_{\beta,z}}.
\eea
Hence, we note that the sign factor $(-)^{i}$ in the second term in the right hand side of Eq. (\ref{eq:flow_h1_u_Ry_2}) can be replaced by either $(-)^{\delta_{\mu,x}+\delta_{\mu,z}+\delta_{\beta,x}+\delta_{\beta,z}}$ (for the $h_{k+1}^{R_y\beta}$ term), or $(-)^{\delta_{\mu,x}+\delta_{\mu,z}+\delta_{\alpha,x}+\delta_{\alpha,z}}$ (for the $h_{k+1}^{R_y\alpha}$ term),
since in whichever case the sign factor is 
$(-)^{\delta_{\alpha,x}+\delta_{\alpha,z}+\delta_{\beta,x}+\delta_{\beta,z}}$
 because of the Kronecker delta $\delta_{\mu\alpha}$ (for the $h_{k+1}^{R_y\beta}$ term), and $\delta_{\mu\beta}$ (for the  $h_{k+1}^{R_y\alpha}$ term).
 
 On the other hand, using $R_y(S^x,S^y,S^z)\rightarrow (-S^x,S^y,-S^z)$,
it can be seen that the invariance of the RG flow equations under the symmetry operation $R_yI$ exactly requires Eq. (\ref{eq:flow_h1_u_Ry_3}).
Hence we conclude that the flow equations have the symmetry imposed by $R_yI$.

\section{Numerical determination for the signs of the ``D" coefficients}  
\label{app:numerics_sign_D}

In this appendix, we study the signs of the five ``D" coefficients.
As in Sec. \ref{subsec:compare_numerics}, we work in the four-sublattice rotated frame and take the parameters as $K+2J=1$, $J=-1$, $\Gamma=0.35$.
DMRG numerical simulations are performed on a system of $L=144$ sites with periodic boundary conditions.
The bond dimension $m$ and truncation error $\epsilon$ in DMRG simulations are taken as $m=1400$ and $\epsilon=10^{-9}$.

We add a small uniform magnetic field along the $z$-direction as
\bea
-h^z_0 \sum_n (S_{1+4n}^z+S_{2+4n}^z+S_{3+4n}^z+S_{4+4n}^z).
\label{eq:h_z0}
\eea
The low energy Hamiltonian can be derived as $-h^z_0 i_D \int dx J^z$.
Using the nonsymmorphic bosonization formulas, the spin expectation values are expected to be
\bea
\langle \vec{S}_{1+4n} \rangle &=& \langle J^z \rangle(h_D,-h_D,i_D),\nn\\
\langle \vec{S}_{2+4n} \rangle &=& \langle J^z \rangle(-h_D,-h_D,i_D),\nn\\
\langle \vec{S}_{3+4n} \rangle &=& \langle J^z \rangle(-h_D,h_D,i_D),\nn\\
\langle \vec{S}_{4+4n} \rangle &=& \langle J^z \rangle(h_D,h_D,i_D).
\label{eq:pattern_hz0_D}
\eea
Choosing $h_0^z=10^{-3}$, DMRG numerical simulations give
\begin{align}
    S^x &\simeq -1.004*10^{-6}*(+,-,-,+)\\
    S^y &\simeq -1.004*10^{-6}*(-,-,+,+)\\
    S^z &\simeq 5.25*10^{-7}*(+,+,+,+).
\end{align}
Comparing with Eq. (\ref{eq:pattern_hz0_D}), we obtain
\bea
\langle J^z\rangle h_D&=&-1.004\times 10^{-6},\nn\\
\langle J^z\rangle i_D&=&5.25\times 10^{-7}.
\label{eq:coeff_value_hz0_D}
\eea

If we add the following mixture of the  magnetic fields in the $xy$-plane,
\bea
-h^{xy}\big[ a_C\sum_n (S_{1+4n}^y+S_{2+4n}^y+S_{3+4n}^y+S_{4+4n}^y)+b_C\sum_n (-S_{1+4n}^x+S_{2+4n}^x-S_{3+4n}^x+S_{4+4n}^x)  \big].
\eea
then the low energy Hamiltonian  is 
\bea
-h^{xy} (a_Ca_D-b_Cb_D) \int dx J^y.
\eea
Using the nonsymmorphic bosonization formulas, the spin expectation values are expected to be
\bea
\langle \vec{S}_{1+4n} \rangle &=& \langle J^y \rangle(b_D,a_D,-c_D),\nn\\
\langle \vec{S}_{2+4n} \rangle &=& \langle J^y \rangle(-b_D,a_D,-c_D),\nn\\
\langle \vec{S}_{3+4n} \rangle &=& \langle J^y \rangle(b_D,a_D,c_D),\nn\\
\langle \vec{S}_{4+4n} \rangle &=& \langle J^y \rangle(-b_D,a_D,c_D).
\eea
Choosing $h^{xy}=10^{-3}$, DMRG numerical simulations give
\begin{align}
    S^x &\simeq -9.53*10^{-5}*(+,-,+,-)\\
    S^y &\simeq 1.30*10^{-6}*(+,+,+,+)\\
    S^z &\simeq -4.13*10^{-6}*(-,-,+,+).
    \label{eq:coeff_value_hxy0_D}
\end{align}

From Eq. (\ref{eq:coeff_value_hz0_D}) and Eq. (\ref{eq:coeff_value_hxy0_D}), the ratios can be determined as 
\bea
h_D/i_D&=&-1.91,\nn\\
b_D/a_D&=& -73.3,\nn\\
c_D/a_D&=&-3.18.
\label{eq:hxy_results}
\eea
Notice that in contrast with Table \ref{table:coeff} where $i_D$ and $a_D$ are the dominant coefficients (which is consistent with RG predictions), 
the absolute values of the ratios in Eq. (\ref{eq:hxy_results})  severely violate the relations in Eq. (\ref{eq:hierarchy}).
The huge discrepancies in the ``D" coefficients when magnetic field responses are studied remain puzzling, and the reasons are unclear.

\end{widetext}



\begin{thebibliography}{10}


\bibitem{Jackeli2009}
G. Jackeli and G. Khaliullin, Phys. Rev. Lett. {\bf 102}, 017205 (2009).

\bibitem{Chaloupka2010}
J. Chaloupka, G. Jackeli, and G. Khaliullin, Phys. Rev. Lett. {\bf 105}, 027204 (2010).

\bibitem{Singh2010}
Y. Singh and P. Gegenwart, Phys. Rev. B {\bf 82}, 064412 (2010).

\bibitem{Price2012}
C. C. Price and N. B. Perkins, Phys. Rev. Lett. {\bf 109}, 187201 (2012).

\bibitem{Singh2012}
Y. Singh, S. Manni, J. Reuther, T. Berlijn, R. Thomale, W. Ku, S. Trebst, and P. Gegenwart, Phys. Rev. Lett. {\bf 108}, 127203 (2012).

\bibitem{Plumb2014}
K. W. Plumb, J. P. Clancy, L. J. Sandilands, V. V. Shankar, Y. F. Hu, K. S. Burch, H. Y. Kee, and Y. J. Kim, Phys. Rev. B {\bf 90}, 041112 (2014).

\bibitem{Kim2015}
H.-S. Kim, V. S. V., A. Catuneanu, and H.-Y. Kee, Phys. Rev. B {\bf 91}, 241110 (2015).

\bibitem{Winter2016}
S. M. Winter, Y. Li, H. O. Jeschke, and R. Valenti, Phys. Rev. B {\bf 93}, 214431 (2016).

\bibitem{Baek2017}
S. H. Baek, S. H. Do, K. Y. Choi, Y. Kwon, A. Wolter, S. Nishimoto, J. van den Brink, and B. Buchner, Phys. Rev. Lett. {\bf 119}, 037201 (2017).

\bibitem{Leahy2017}
I. A. Leahy, C. A. Pocs, P. E. Siegfried, D. Graf, S. H. Do, K. Y. Choi, B. Normand, and M. Lee, Phys. Rev. Lett. {\bf 118}, 187203 (2017).

\bibitem{Sears2017}
J. A. Sears, Y. Zhao, Z. Xu, J. W. Lynn, and Y. J. Kim, Phys. Rev. B {\bf 95}, 180411 (2017).

\bibitem{Wolter2017}
A. U. B. Wolter, L. T. Corredor, L. Janssen, K. Nenkov, S. Schonecker, S. H. Do, K. Y. Choi, R. Albrecht, J. Hunger, T. Doert, M. Vojta, and B. Buchner, Phys. Rev. B {\bf 96}, 041405(R) (2017).

\bibitem{Zheng2017}
J. Zheng, K. Ran, T. Li, J. Wang, P. Wang, B. Liu, Z.-X. Liu, B. Normand, J. Wen, and W. Yu, Phys. Rev. Lett. {\bf 119}, 227208 (2017).

\bibitem{Rousochatzakis2017} 
I. Rousochatzakis and N. B. Perkins, Phys. Rev. Lett. {\bf 118}, 147204 (2017).

\bibitem{Kasahara2018}
Y. Kasahara, T. Ohnishi, N. Kurita, H. Tanaka, J. Nasu, Y. Motome, T. Shibauchi, and Y. Matsuda, Nature (London) {\bf 559}, 227 (2018).

\bibitem{Rau2014}
J. G. Rau, E. K. H. Lee, and H. Y. Kee, Phys. Rev. Lett. {\bf 112}, 077204 (2014).

\bibitem{Ran2017}
K. Ran, J. Wang, W. Wang, Z.-Y. Dong, X. Ren, S. Bao, S. Li, Z. Ma, Y. Gan, Y. Zhang, J. T.  Park, G. Deng, S. Danilkin, S.-L. Yu, J.-X. Li, and J. Wen, Phys. Rev. Lett. {\bf 118}, 107203 (2017).

\bibitem{Wang2017}
W. Wang, Z.-Y. Dong, S.-L. Yu, and J.-X. Li, Phys. Rev. B {\bf 96}, 115103 (2017).

\bibitem{Catuneanu2018}
A. Catuneanu, Y. Yamaji, G. Wachtel, Y. B. Kim, and H.-Y. Kee, npj Quantum Mater. {\bf 3}, 23 (2018).

\bibitem{Gohlke2018}
M. Gohlke, G. Wachtel, Y. Yamaji, F. Pollmann, and Y. B. Kim, Phys. Rev. B {\bf 97}, 075126 (2018).

\bibitem{Liu2011}
 X. Liu, T. Berlijn, W.-G. Yin, W. Ku, A. Tsvelik, Young-June Kim, H. Gretarsson, Yogesh Singh, P. Gegenwart, and J. P. Hill,
 Phys. Rev. B \textbf{83}, 220403(R) (2011).

\bibitem{Chaloupka2013}
J. Chaloupka, G. Jackeli, and G. Khaliullin,
Phys. Rev. Lett. \textbf{110}, 097204 (2013).

\bibitem{Johnson2015}
R. D. Johnson, S. C. Williams, A. A. Haghighirad, J. Singleton, V. Zapf, P. Manuel,
I. I. Mazin, Y. Li, H. O. Jeschke, R. Valent\'i, and R. Coldea,
Phys. Rev. B \textbf{92}, 235119 (2015).

\bibitem{Motome2020}
Motome, R. Sano, S. H. Jang, Y. Sugita, and Y. Kato, J. Phys.: Condens. Matter 32, 404001 (2020).

\bibitem{Kitaev2006}
A. Kitaev, Ann. Phys. (N. Y). {\bf 321}, 2 (2006).

\bibitem{Nayak2008}
C. Nayak, S. H. Simon, A. Stern, M. Freedman, and S. Das Sarma, Rev. Mod. Phys. {\bf 80}, 1083 (2008).


\bibitem{Fazekas1999}
P. Fazekas, {\it Lecture Notes on Electron Correlation and Magnetism}, Vol. 5 of {\it Series in Modern Condensed Matter Physics} (World Scientific, 1999).

\bibitem{Lauchli2006}
A. L{\"a}uchli, F. Mila, and K. Penc, Phys. Rev. Lett. {\bf 97}, 087205 (2006).


\bibitem{Balents2010}
L. Balents, Nature {\bf 464}, 199 (2010).

\bibitem{Witczak-Krempa2014}
W. Witczak-Krempa, G. Chen, Y. B. Kim, and L. Ba-
lents, Annu. Rev. Condens. Matter Phys. {\bf 5}, 57 (2014).

\bibitem{Rau2016}
J. G. Rau, E. K.-H. Lee, and H.-Y. Kee, Annu. Rev.
Condens. Matter Phys. {\bf 7}, 195 (2016).

\bibitem{Winter2017}
S. M. Winter, A. A. Tsirlin, M. Daghofer, J. van den
Brink, Y. Singh, P. Gegenwart, and R. Valenti, J. Phys. Condens. Matter {\bf 29}, 493002 (2017).

\bibitem{Zhou2017}
Y. Zhou, K. Kanoda, and T. K. Ng, Rev. Mod. Phys. {\bf 89}, 025003 (2017).

\bibitem{Savary2017}
L. Savary, L. Balents, Rep. Prog. Phys. {\bf 80}, 016502 (2017).


\bibitem{Sela2014}
E. Sela, H.-C. Jiang, M. H. Gerlach, and S. Trebst, Phys. Rev. B {\bf 90}, 035113 (2014).

\bibitem{Agrapidis2018}
C. E. Agrapidis, J. van den Brink, and S. Nishimoto, Sci. Rep. {\bf 8}, 1815 (2018).

\bibitem{Agrapidis2019}
C. E. Agrapidis, J. van den Brink, and S. Nishimoto, Phys. Rev. B {\bf 99}, 224418 (2019).

\bibitem{Catuneanu2019}
A. Catuneanu, E. S. S\o rensen, and H.-Y. Kee, Phys. Rev. B {\bf 99}, 195112 (2019).

\bibitem{Yang2020}
W. Yang, A. Nocera, T. Tummuru, H.-Y. Kee, and I. Affleck, Phys. Rev. Lett. {\bf 124}, 147205 (2020).

\bibitem{Yang2020a}
W. Yang, A. Nocera, and I. Affleck, Phys. Rev. Research {\bf 2}, 033268 (2020).

\bibitem{Yang2020b}
W. Yang, A. Nocera, and I. Affleck, Phys. Rev. B {\bf 102}, 134419 (2020).

\bibitem{Yang2021b}
W. Yang, A. Nocera, E. S. S\o rensen, H.-Y. Kee, and I. Affleck,  Phys. Rev. B {\bf 103}, 054437 (2021).

\bibitem{Yang2022}
W. Yang, A. Nocera, P. Herringer, R. Raussendorf, I. Affleck, Phys. Rev. B {\bf 105}, 094432 (2022).

\bibitem{Yang2022b}
W. Yang, C. Xu, S. Xu, A. Nocera, I. Affleck, arXiv:2202.11686 (2022).

\bibitem{Luo2021}
Q. Luo, J. Zhao, X. Wang, and H.-Y. Kee, Phys. Rev. B {\bf 103}, 144423(2021).

\bibitem{Luo2021b}
Q. Luo, S. Hu, and H.-Y. Kee, Phys. Rev. Research {\bf 3}, 033048 (2021).

\bibitem{You2020}
Z.-A. Liu, T.-C. Yi, J.-H. Sun, Y.-L. Dong, and W.-L. You, Phys. Rev. E {\bf 102}, 032127 (2020).

\bibitem{Sorensen2021}
E. S. S\o rensen, A. Catuneanu, J. Gordon, H.-Y. Kee, Phys. Rev. X {\bf 11}, 011013 (2021). 


\bibitem{Affleck1988}
I. Affleck, in {\it Fields, Strings and Critical Phenomena}, Proceedings of Les Houches Summer School, 1988, edited by E. Brezin and J. Zinn-Justin (North-Holland, Amsterdam, 1990), pp. 563-640.

\bibitem{Amit1984}
D.J. Amit, {\it Field Theory, the Renormalization Group and Critical Phenomena}, 2nd Edition, World Scientific, 1984.


\end{thebibliography}
\end{document}